\documentclass[twocolumn]{autart}     

\usepackage{graphicx}          

\usepackage{amsmath,amssymb,amsfonts}
\usepackage{setspace}
\usepackage{cite}
\usepackage{graphicx}
\usepackage{textcomp}
\usepackage{xcolor}
\usepackage{lipsum}
\usepackage{arydshln}
\usepackage{mwe}
\def\BibTeX{{\rm B\kern-.05em{\sc i\kern-.025em b}\kern-.08em
    T\kern-.1667em\lower.7ex\hbox{E}\kern-.125emX}}
    
\usepackage{mathtools}
\usepackage{cuted}
\usepackage{dsfont}

\usepackage{optidef}

\usepackage{pgf,tikz}
\usetikzlibrary{positioning}
\usepackage{pgfplots}
\pgfplotsset{compat = newest}
\pgfplotsset{every x tick label/.append style={font=\footnotesize, yshift=0.5ex}}
\pgfplotsset{every y tick label/.append style={font=\footnotesize, xshift=0.5ex}}

\usepackage{url}

\newcommand{\sign}{\mathop{\bf sign}}

\newcommand{\ie}{{\it i.e.\;}}

\DeclarePairedDelimiter\parentheses{\lparen}{\rparen}

\newcommand{\rank}[1]{\operatorname{rank} \parentheses*{#1}}
\newcommand{\spanv}[1]{\operatorname{span} \parentheses* {#1}}
\newcommand{\trace}[1]{\operatorname{trace} \parentheses*{#1}}

\newcommand{\colspan}[1]{\operatorname{colspace}\left(#1\right)\,}

\newcommand{\kernel}[1]{\operatorname{kernel}\left(#1\right)\,}

\newcommand{\target}{\mathcal{T}}

\tikzstyle{neuron} = [draw, fill=white, circle, 
    minimum height=3.5em, minimum width=1em, style=thick]

\newcommand{\K}{\mathbf{K}}

\newtheorem{theorem}{Theorem}
\newtheorem{proposition}{Proposition}

\newtheorem{corollary}{Corollary}

\newtheorem{definition}{Definition}
\newtheorem{lemma}{Lemma}

\newtheorem{remark}{Remark}
\DeclareMathOperator{\tr}{\mathrm{trace}}
\allowdisplaybreaks

\definecolor{darkgreen}{HTML}{008000}
\definecolor{darkred}{HTML}{A52A2A}

\newcommand{\comment}[1]{}
\newcommand{\edo}{\end{document}}

\newcommand{\R}{{\mathbb R}}  
\newcommand{\Z}{{\mathbb Z}}  

\newcommand{\be}[1]{\begin{equation}\label{#1}}
\newcommand{\ee}{\end{equation}}
\newcommand{\halmos}{\rule{1ex}{1.4ex}}
\newcommand{\beq}{\begin{eqnarray}}
\newcommand{\eeq}{\end{eqnarray}}
\newcommand{\beqn}{\begin{eqnarray*}}
\newcommand{\eeqn}{\end{eqnarray*}}
\newcommand{\bi}{\begin{itemize}}
\newcommand{\ei}{\end{itemize}}
\newcommand{\ben}{\begin{enumerate}}
\newcommand{\een}{\end{enumerate}}
\newcommand{\bes}[1]{\begin{subequations}\label{#1}\begin{eqnarray}}
\newcommand{\ees}[1]{\end{eqnarray}\end{subequations}}

\newcommand{\equilx}{\bar{x}}
\newcommand{\equilset}{E}
\newcommand{\initialx}{\xi }
\newcommand{\statespace}{\mathbb{X}}
\newcommand{\domain}{\mathcal{D}_{\mbox{\tiny F}}}

\newcommand{\Tmax}{T^{\mbox{\tiny max}}_{\initialx}}
\newcommand{\Tmin}{T^{\mbox{\tiny min}}_{\initialx}}
\newcommand{\Thit}{\tau _{\initialx}}
\newcommand{\jac}[1]{J_{#1}}
\newcommand{\convergingset}{{\mathcal C}_{\equilset}}
\newcommand{\loss}{\mathcal{L}}

\newcommand{\circled}[1]{\raisebox{.5pt}{\textcircled{\raisebox{-0.0pt} {\tiny #1}}}}

\newcommand{\NNReals}{\mathbb{R}_+}
\newcommand{\PosReals}{\mathbb{R}_{++}}
\newcommand{\Reals}[1][]{\mathbb{R}^{#1}}
\newcommand{\PSD}[1][]{\mathbb{S}_+^{#1}}
\newcommand{\PD}[1][]{\mathbb{S}_{++}^{#1}}
\newcommand{\Kstbl}{\mathcal{K}}
\newcommand{\XZ}{\mathcal{X}_0}

\newcommand{\Kb}{\overline{\mathbf{K}}}
\newcommand{\Ks}{\mathbf{K^*}}

\newcommand{\Ccal}[1][]{\mathcal{C}_{#1}}
\newcommand{\dd}{\mathrm{d}}
\renewcommand{\qed}{\hfill \halmos} 
    \usepackage[normalem]{ulem}  
    \colorlet{highlight}{red!70!black}
    \newcommand{\old}[1]{#1}

    \renewcommand{\sout}[1]{}
    \renewcommand{\old}[1]{}
    \colorlet{highlight}{red!00!black}

\newcommand{\highlight}[1]{{\color{highlight} #1}}

\allowdisplaybreaks

\begin{document}

\begin{frontmatter}
\runtitle{Overparameterized LQR Formulations}  

\title{Convergence Analysis of Gradient Flow for Overparameterized LQR Formulations\thanksref{footnoteinfo}} 

\thanks[footnoteinfo]{\color{red!00!black}This work was partially supported by ONR Grant N00014-21-1-2431,  NSF Grant 2121121, and AFOSR Grant FA9550-21-1-0289.}

\author{Arthur Castello B. de Oliveira}\ead{castello.a@northeastern.edu},    
\author{Milad Siami}\ead{m.siami@northeastern.edu},               
\author{Eduardo D. Sontag}\ead{e.sontag@northeastern.edu}  




\begin{abstract}                          
    This paper analyses the intersection between results from gradient methods for the model-free linear quadratic regulator (LQR) problem, and linear feedforward neural networks (LFFNNs). More specifically, it looks into the case where one wants to find a LFFNN feedback that minimizes a LQR cost. It starts by deriving a key conservation law of the system, which is then leveraged to generalize existing results on boundedness and global convergence of solutions, and invariance of the set of stabilizing LFFNNs under the training dynamics (gradient flow). For the single hidden layer LFFNN, the paper proves that the ``training'' converges to the optimal feedback control law for all but a set of Lebesgue measure zero of the initializations. These results are followed by an analysis of a simple version of the problem -- the ``vector case'' -- proving the theoretical properties of accelerated convergence and a type of input-to-state stability (ISS) result for this simpler example. Finally, the paper presents numerical evidence of faster convergence of the gradient flow of general LFFNNs when compared to non-overparameterized formulations, showing that the acceleration of the solution is observable even when the gradient is not explicitly computed, but estimated from evaluations of the cost function.
\end{abstract}

\end{frontmatter}

\section{Introduction}

Neural networks and machine learning (ML) tools are being increasingly used in control design \cite{wafi2023comparative, alali2023reinforcement, zandi2023robot, motamedi2024robustness, ravari2022optimal, ravari2024optimal,sznaier2022role}, and are particularly useful in model-free applications, where a model of the system might not be available \cite{fazel2018global,cui2024small}. In such scenarios, an ``oracle'' might be queried to estimate the cost associated with a specific control law, as illustrated in Fig. \ref{fig:uncertainoracle}. This feedback has adjustable parameters (or ``weights''), which are updated through the gradient of the estimated cost, typically employing gradient descent or some other similar numerical optimization method.

\begin{figure}[t]
    \centering
    {\begin{tikzpicture}[scale=0.9,transform shape,node distance=6.5cm]
        \node[rectangle, rounded corners, name=Algorithm, style=very thick, draw, align=center,minimum size=1.5cm] at (0,0) {\Large Design \\ \vspace{-7pt}\\ \Large Algorithm};
        \node [rectangle, rounded corners, name=Oracle, right of = Algorithm, style=very thick, draw, align=center,minimum size=1.5cm] {\Large Uncertain \\ \vspace{-7pt}\\ \Large Oracle};

        \path (Algorithm.east) -- (Algorithm.north east) coordinate[pos=0.5] (a1);
        \path (Oracle.west) -- (Oracle.north west) coordinate[pos=0.5] (o1);
        \draw[latex-, line width=0.5mm] (a1) -- node [above] {\tiny \bf Cost Evaluation} (o1);

        \path (Algorithm.east) -- (Algorithm.south east) coordinate[pos=0.5] (a2);
        \path (Oracle.west) -- (Oracle.south west) coordinate[pos=0.5] (o2);
        \draw[-latex, line width=0.5mm] (a2) -- node [below] {\tiny \bf Current Feedback Matrix} (o2);
    \end{tikzpicture}}
    \caption{\small System overview of the model-free control design process. The design algorithm attempts to find a feedback matrix that minimizes the output of the oracle, which in turn provides a (possibly noisy) estimate of the cost function every time it receives a candidate feedback matrix.}
    \label{fig:uncertainoracle}
\end{figure}

Understanding the convergence of such learning techniques is challenging due to inherent nonlinearities. In particular, neural networks leverage both their compositional structure and the nonlinear activation functions of each layer. Previous works on neural networks isolate the effects of its compositional structure from the nonlinear activation by studying linear feedforward neural networks (LFFNNs) \cite{bah_learning_2022,chitour_geometric_2023,de2023dynamics,de2024remarks,eftekhari_training_2020,kawaguchi_deep_2016,min_explicit_2021,min2023convergence}. The results are typically given for solving a static supervised learning problem, \ie a linear regression of labels $u$ on $K_N\dots K_1 y$, where $y$ is an input. Not only powerful ``almost everywhere'' convergence results have been obtained for the regression problem \cite{kawaguchi_deep_2016, bah_learning_2022, baldi_neural_1989, eftekhari_training_2020, chitour_geometric_2023, arora2019fine}, and the a type of input-to-state stability (ISS) property of an associated problem was characterized \cite{de2023dynamics}, but, perhaps surprisingly, the optimization on the individual matrices $K_i$ can result in much faster convergence than optimization on a single matrix $K$ \cite{min_explicit_2021, min2023convergence, tarmoun_understanding_2021}.

Despite the rich literature, current results on LFFNNs cannot be applied out-of-the-box to non-convex problems, even if under some gradient dominance condition (PL-inequality). An extremely popular and well-studied example of such a system is the linear quadratic regulator (LQR) problem, whose general goal is to minimize a quadratic cost function $J(K)$, where $K$ is a candidate feedback law. When the system dynamics are linear and known, an explicit optimal solution is obtainable by solving a Riccati equation on the system matrices \cite{sontag2013mathematical}, but such an approach is generally unfit for model-free scenarios, where the system is assumed unknown and only the value of the cost function for different feedback matrices can be queried to some oracle (as illustrated in Fig. \ref{fig:uncertainoracle}). This type of scenario can be understood as a ``policy optimization'' formulation for the LQR problem (poLQR) \cite{hu2023toward}, and approximates the LQR problem to reinforcement learning problems, motivating previous works where the optimization is solved by following the negative flow of the gradient $\dot K = -\nabla J(K)$, or negative descent direction $K_{n+1}=K_n-h\nabla J(K_n)$ (for some step-size $h>0$) \cite{levine_determination_1970,fazel2018global,mohammadi_convergence_2022,hu2023toward,sontag_remarks_2022,cui2024small}. Such ``training'' is an area of active research to this day due to its non-convex landscape, and traces its origins to pioneering work by Levine and Athans starting in the late 1960s \cite{levine_determination_1970}. Recent publications have established global convergence properties \cite{fazel2018global,mohammadi_convergence_2022,hu2023toward}, as well as input-to-state stability (ISS) \cite{sontag_remarks_2022,cui2024small} when the computation of the gradient is subject to error or uncertainty. Of special note, in \cite{hu2023toward}, the authors explore the relationship between LQR (and other classical control problems) and policy optimization, leveraging the explicit expression of the gradient of the linear quadratic cost to prove the convergence of gradient descent methods. Similarly, in \cite{zhang2023revisiting}, the authors also interpret a gradient approach to this problem as policy optimization for the LQR problem, and explore the relationship between the finite and infinite horizon formulations of the LQR problem to propose a strategy that converges even if initialized outside the set of stabilizing controllers. All these results argue for the importance of understanding the behavior of the gradient flow when studying the LQR problem in a model-free context.

In this context, the primary goal of this paper is to study the effects of LFFNNs when applied to the more complex setting of solving a model-free LQR problem (poLQR). Mathematically, the feedback is written as a product $K=K_N\dots K_1$, where $K_i$ represents the weights of the $i$th layer of the network. In this context, the natural training dynamics take the form $\dot K_i=-\nabla_{K_i}J(K_N\dots K_1)$ for $i=1,\dots, N$, which is a coupled set of gradient flows done on the full set of parameters $(K_1,\dots,K_N)$. In particular, the assumptions of gradient dominance (PL inequality) and coerciveness of the cost function -- important for both general non-convex optimization \cite{polyak1963gradient,jin2017escape,agarwal2021theory,nesterov2006cubic} and for gradient methods for solving the poLQR \cite{fazel2018global,gravell2020learning,mohammadi2021lack} -- do not hold when optimizing over layers of a LFFNN. This is due to the introduction of spurious equilibria and multiple non-compact sets of critical points in the gradient dynamics. Despite those issues, we derive convergence properties of the solution of an overparameterized formulation for the poLQR. 

Beyond the original goal, literature results on accelerated convergence \cite{min_explicit_2021,min2023convergence,tarmoun_understanding_2021} indicate that even the simpler problem of linear activation functions can be interesting and useful from more than just a theoretical point of view. We demonstrate that this property also holds for the poLQR through numerical simulations, although further discussions on computational and sample complexity are required before it can be determined whether this formulation is inherently useful for practical applications.

In sum, this paper takes steps to blend these two strands of research: gradient methods for the model-free LQR problem; and the analysis of overparameterization in optimization. It looks at the use of overparameterized state feedback for the poLQR, investigating properties that can be derived for its gradient flow.

To accomplish this, the paper starts at Section \ref{sec:ThrBknd} presenting a theoretical background of both gradient methods for the LQR problem and overparameterization for linear regression problems. Then, in Section \ref{sec:NNLQR} the paper formally defines the overparameterized formulation for the LQR problem, and it proves that it shares the same convergence properties as the overparameterized linear regression. Then, Section \ref{sec:scalarcase} does a complete characterization of a simplified version of the problems: the case of single input and single state/output. In this section, the center-stable manifold of the spurious equilibria is characterized and both a type of ISS and a accelerated convergence properties are formally proven. The paper then presents numerical simulations to demonstrate the presence of accelerated convergence for the general case in Section \ref{sec:simul}. The simulations show how initialization affects convergence when compared to the non-overparameterized gradient flow, both when the gradient is perfectly and imperfectly known. Finally, in Section \ref{sec:conclusions} the contributions of this paper are summarized and possible future directions of work are discussed. A preliminary version of this work was previously published \cite{de2024remarks}, however, the proofs appear here for the first time, and the discussion is significantly deepened. Furthermore, additional auxiliary results and a more thorough set of simulations are presented only in this version, providing a much clearer intuition of the behavior of the system. All proofs are provided in the appendix for the clarity of the main text.

\section{Theoretical background}
\label{sec:ThrBknd}


Along this paper, let $\NNReals$ and $\PosReals$ be the set of nonnegative and strictly positive real numbers respectively. For $n\in\mathbb{N}$, let $\PSD[n]$ and $\PD[n]$ be the set of symmetric positive semi-definite (PSD) and positive definite (PD) $n$-by-$n$ matrices, respectively. Given a matrix $A\in\Reals[n\times n]$, $A$ is said to be Hurwitz if all its eigenvalues have negative real part. 

\subsection{The LQR problem as policy optimization}

We begin by presenting results from \cite{rautert_computational_1997}, which serve as groundwork upon which we derive our new results. We also emphasize that despite the reliance of the following results on the knowledge of the system matrices, the gradient expression derived in this section holds great value for analysis, as demonstrated, for example, in \cite{fazel2018global, mohammadi_convergence_2022,hu2023toward}, where it forms the basis for theoretical guarantees regarding convergence rate and accuracy in model-free scenarios.

Consider the following linear system:
\begin{equation}
    \label{eq:ltisys}
    \Sigma\begin{cases}
        \dot x = Ax+Bu \\ y = Cx
    \end{cases},
\end{equation}
where $A\in\Reals[n\times n]$, $B\in\Reals[n\times m]$, and $C\in\Reals[n\times n]$ are the system matrices, with $(A,B)$ assumed controllable and $C$ assumed full rank (Assumption 2 of \cite{rautert_computational_1997}). Let $\Kstbl:=\{K\in\Reals[m\times n] ~|~ A+BKC \text{ is Hurwitz}\}$, then the objective is to determine an output feedback $u=Ky$, $K\in\Kstbl$, that minimizes
\begin{equation}
    \label{eq:JLQR}
    J(K) = \mathbb{E}_{x_0\sim\XZ}\left[\int_0^\infty x(t)^\top Q x(t) + u(t)^\top R u(t) \, dt\right],
\end{equation}
with given positive definite cost matrices $R\in\PD[m\times m]$ and $Q\in\PD[n\times n]$, and for $x_0$ sampled from a probability distribution $\XZ$. Along this paper we refer to $K^*$ as the unique solution to the LQR problem.

In \cite{rautert_computational_1997}, the authors provide, through Theorem 3.2, the following expression for the gradient $\nabla J$ with respect to the feedback matrix $K$:
\begin{equation}
    \label{eq:gradJLQR}
    \nabla J(K) = 2(B^\top P_K + RKC) L_KC^\top,
\end{equation}
where for any $K\in\Kstbl$, $P_K$ and $L_K$ are the unique positive definite solutions of the following Lyapunov equations
\begin{align}
    P_K(A+BKC) + (A+BKC)^\top P_K &\nonumber \\+ C^\top K^\top RKC + Q &= 0 \label{eq:LyapPtrad}\\
    L_K(A+BKC)^\top + (A+BKC)L_K + \Sigma_0 &= 0, \label{eq:LyapLtrad}
\end{align}
respectively, and the matrix $\Sigma_0=\mathbb{E}_{x_0\sim \XZ}[x_0x_0^\top]$ depends on the distribution of initial conditions $\XZ$, and is assumed to be full rank. 
From these, we can define the set of desired/optimal value of $K$ as $\target := \{K\in\Kstbl ~|~ \nabla J(K)=0\}$.
With these results established, we next look at key literature results on overparameterization.

\subsection{Overparameterization - properties and formulation}

\begin{figure}
    \centering
    {\begin{tikzpicture}[scale=0.75,transform shape,node distance=1.3cm]
        \node [neuron, name=neui1] { $y_1$};
        \node [rectangle, name=neui2, below of = neui1] { \vdots};
        \node [neuron, name=neui3, below of = neui2] { $y_n$};
        
        \node [rectangle, name=neuh13, right of = neui2, node distance = 2.5cm] { \vdots};
        \node [neuron, name=neuh12, above of = neuh13] { $z^1_2$};
        \node [neuron, name=neuh14, below of = neuh13] { $z^1_{\kappa_1}$};

        \node [rectangle, name=neud1, right of = neuh13, node distance = 2.5cm] {\bf$\ddots$};
        \node [rectangle, name=neud2, above of = neud1] {\bf$\ddots$};
        \node [rectangle, name=neud3, below of = neud1] {\bf$\ddots$};

        \node [rectangle, name=neuhN3, right of = neud1, node distance = 2.5cm] { \vdots};
        \node [neuron, name=neuhN2, above of = neuhN3] { $z^{N-1}_1$};
        \node [neuron, name=neuhN1, below of = neuhN3] { $z^{N-1}_{\kappa_{N -1}}$};

        \node [rectangle, name=neuy1, right of = neuhN3, node distance = 2.5cm] {\vdots};
        
        \node [neuron, name=neuy2, above of = neuy1] { $u_1$};
        \node [neuron, name=neuy3, below of = neuy1] { $u_m$};

        \node [rectangle, name=aux1, right of = neui1, node distance = 1.25cm] {};
        \node [rectangle, name=K1, above of = aux1, node distance = 0.5cm] {\color{blue}$K_1$};
        \node [rectangle, name=K2, right of = K1, node distance = 2.5cm] {\color{red!70!black}$K_2$};
        \node [rectangle, name=KN1, right of = K2, node distance = 2.5cm] {\color{green!60!black}$K_{N-1}$};
        \node [rectangle, name=KN, right of = KN1, node distance = 2.5cm] {\color{yellow!60!black}$K_N$};

        \draw [->, line width=0.3mm, color=blue] (neui1) -- (neuh12) {};
        \draw [->, line width=0.3mm, color=blue] (neui1) -- (neuh13) {};
        \draw [->, line width=0.3mm, color=blue] (neui1) -- (neuh14) {};
        
        \draw [->, line width=0.3mm, color=blue] (neui2) -- (neuh12) {};
        \draw [->, line width=0.3mm, color=blue] (neui2) -- (neuh13) {};
        \draw [->, line width=0.3mm, color=blue] (neui2) -- (neuh14) {};
        
        \draw [->, line width=0.3mm, color=blue] (neui3) -- (neuh12) {};
        \draw [->, line width=0.3mm, color=blue] (neui3) -- (neuh13) {};
        \draw [->, line width=0.3mm, color=blue] (neui3) -- (neuh14) {};
        
        \draw [->, line width=0.3mm, color=red!70!black] (neuh12) -- (neud1) {};
        \draw [->, line width=0.3mm, color=red!70!black] (neuh13) -- (neud1) {};
        \draw [->, line width=0.3mm, color=red!70!black] (neuh14) -- (neud1) {};

        \draw [->, line width=0.3mm, color=red!70!black] (neuh12) -- (neud2) {};
        \draw [->, line width=0.3mm, color=red!70!black] (neuh13) -- (neud2) {};
        \draw [->, line width=0.3mm, color=red!70!black] (neuh14) -- (neud2) {};

        \draw [->, line width=0.3mm, color=red!70!black] (neuh12) -- (neud3) {};
        \draw [->, line width=0.3mm, color=red!70!black] (neuh13) -- (neud3) {};
        \draw [->, line width=0.3mm, color=red!70!black] (neuh14) -- (neud3) {};

        \draw [->, line width=0.3mm, color=green!60!black] (neud1) -- (neuhN1) {};
        \draw [->, line width=0.3mm, color=green!60!black] (neud1) -- (neuhN2) {};
        \draw [->, line width=0.3mm, color=green!60!black] (neud1) -- (neuhN3) {};

        \draw [->, line width=0.3mm, color=green!60!black] (neud2) -- (neuhN1) {};
        \draw [->, line width=0.3mm, color=green!60!black] (neud2) -- (neuhN2) {};
        \draw [->, line width=0.3mm, color=green!60!black] (neud2) -- (neuhN3) {};

        \draw [->, line width=0.3mm, color=green!60!black] (neud3) -- (neuhN1) {};
        \draw [->, line width=0.3mm, color=green!60!black] (neud3) -- (neuhN2) {};
        \draw [->, line width=0.3mm, color=green!60!black] (neud3) -- (neuhN3) {};

         \draw [->, line width=0.3mm, color=yellow!60!black] (neuhN1) -- (neuy1) {};
        \draw [->, line width=0.3mm, color=yellow!60!black] (neuhN2) -- (neuy1) {};
        \draw [->, line width=0.3mm, color=yellow!60!black] (neuhN3) -- (neuy1) {};

        \draw [->, line width=0.3mm, color=yellow!60!black] (neuhN1) -- (neuy2) {};
        \draw [->, line width=0.3mm, color=yellow!60!black] (neuhN2) -- (neuy2) {};
        \draw [->, line width=0.3mm, color=yellow!60!black] (neuhN3) -- (neuy2) {};

        \draw [->, line width=0.3mm, color=yellow!60!black] (neuhN1) -- (neuy3) {};
        \draw [->, line width=0.3mm, color=yellow!60!black] (neuhN2) -- (neuy3) {};
        \draw [->, line width=0.3mm, color=yellow!60!black] (neuhN3) -- (neuy3) {};
    \end{tikzpicture}}
    \caption{\small Graphical representation of a linear feedforward neural network (LFFNN) with an input layer $y\in\mathbb{R}^n$ with $n$ neurons, hidden layers $z^i\in\mathbb{R}^{\kappa_i}$ with $\kappa_i$ neurons, and output layer $u\in\mathbb{R}^m$ with $m$ neurons. The computation of the network is done for each layer as $z^i = K_iz_{i-1}$, with $z^0=y$ and $z^N=u$, where the matrices $K_i$ represent, in the figure, the presence and weight of edges between neurons of layer $i-1$ and layer $i$. The resulting input-output expression for the LFFNN then becomes $u=K_N\dots K_1y$.}
    \label{fig:LinFFNN}
\end{figure}

The optimization landscape of the gradient flow of a linear neural networks is usually studied in terms of least square/linear regression problems, stated as follows: let $Y=[y_1,y_2,\dots,y_k]$ and $U=[u_1,u_2,\dots, u_k]$ be the column concatenation of (possibly noisy) $k$ input-output pairs sampled from an unknown function $\Kb$ that one wants to approximate using a linear neural network $\K$. Although arguably a simple formulation, the resulting gradient system is the object of study of many papers in the literature \cite{bah_learning_2022,chitour_geometric_2023,de2023dynamics,de2024remarks,eftekhari_training_2020,kawaguchi_deep_2016,min_explicit_2021,min2023convergence}. 

For some search space of neural networks $\Kstbl$, defined as appropriate to the problem, a optimal neural network $\Ks\in\Kstbl$ minimizes $J(\K)=\|U-\K(Y)\|$, where $\K(Y) = [\K(y_1),\dots,\K(y_k)]$, and for some norm $\|\cdot \|$. A linear feedforward neural network (LFFNN) (depicted in Fig. \ref{fig:LinFFNN}) is a feedforward neural network with linear activation functions between layers, and has: an input layer with $n$ neurons; $N-1$ hidden layers, each with $\kappa_i\geq\max(m,n)$ neurons, for $i=1,\dots,N-1$; and an output layer with $m$ neurons. Then, in the specific case of a LFFNN, and being $K_i\in\Reals[\kappa_i\times \kappa_{i-1}]$ the $i$th layer parameter matrix, the function to be minimized becomes $J(K_1,\dots,K_N)=\|U-K_N\dots K_1Y\|$.

For this problem, and under some reasonable assumptions on the ranks of $Y$ and $U$, and on the dimensions of the $K_i$s (see \cite{kawaguchi_deep_2016}, and Assumptions 1 and 2 in \cite{chitour_geometric_2023} and references therein, or a previous work from the authors \cite{de2023dynamics}), the following can be summarized from the literature about the optimization landscape of this problem: 

\begin{proposition}
    Consider a linear regression problem solved with a LFFNN with $N$ layers and trained through gradient flow. Assume $U$ and $Y$ are full column rank and that all hidden layers are wider than the number of inputs and outputs (\ie all hidden layers have more neurons than the input and output layers), then:
    \begin{enumerate}
        \item the problem is generally non-convex and non-concave;
        \item all local minima are global minima;
        \item 
        there are no local maxima;     
        \item in the special case where $N=2$, all critical points are either global minima or strict saddles (\ie the Hessian at that point has at least one strictly negative eigenvalue);
        \item the solution exists for any initial condition and always converges to a critical point of the dynamics;
        \item if $N=2$, the solutions converge to a global optimum for all initializations but a set of Lebesgue measure zero.
    \end{enumerate}
\end{proposition}

    {\it Proof.} Items (1) to (4) are studied in \cite{baldi_neural_1989} for the single hidden layer case, and \cite{kawaguchi_deep_2016} generalized these results to the arbitrarily deep case. Properties (5) and (6) are proved in \cite{2017_panageas_et_al_gradient_descent} for the analogous discretized problem (\ie gradient descent). In Appendix \ref{apx:StrictSaddles} we will show how to adapt these proofs to the continuous-time (\ie flow) case. An independent proof of (5) and (6) was provided in \cite{chitour_geometric_2023} for the specific problem of linear regression and under an additional assumption on the loss function (``distinct critical values'').

Furthermore, other works in the literature establish useful properties of overparameterized linear neural networks, when compared to equivalent non-overparameterized formulations. In \cite{tarmoun_understanding_2021, min_explicit_2021} the authors study the speed of convergence of the gradient flow in overparameterized linear neural networks solving linear regressions, showing that depending on the initialization of the algorithm, the convergence rate can be arbitrarily increased. In \cite{min2023convergence} the authors extend their results to a more general class of optimization problems, however, the required assumption of convexity of the non-overparameterized problem makes it so that their results are not immediately applicable to the LQR problem.

In our previous work \cite{de2023dynamics}, we provide some insights on the loss of robustness in training overparameterized linear neural networks through gradient flow, and show how judicious restrictions on the set of initializations might circumvent this problem.

Such properties for linear neural networks/ overparameterized linear regressions could be useful if they held in the context of feedback control design. Motivated by these results, the next section looks at how one can extend these important results for the policy optimization LQR problem, and consequently to feedback control design.

\section{Feedback control through LFFNNs}
\label{sec:NNLQR}

Let $\K=(K_1,K_2,\dots, K_n)$ be a LFFNN with $N-1$ hidden layers, an input layer, and an output layer. Let $K_1,K_2,\dots,K_N$ be the weight matrices of each layer with $K_1\in\Reals[\kappa_1\times n]$, $K_2\in\Reals[\kappa_2\times \kappa_1]$ and so forth, with $K_N\in\Reals[m\times \kappa_{N-1}]$, where $\kappa_i\in\mathbb{Z}^+$ is the dimension of the $i$th hidden layer. Furthermore, we are interested in the overparameterized case, \ie $\kappa_i\geq\max(m,n)$ for all $i=1,\dots,N-1$. For an input $y\in\Reals[n]$ of the LFFNN, its output $u\in\Reals[m]$ is given by $u = \K(y) = K_N K_{N-1}\cdots K_2K_1 y$,
and its structure is as depicted in Fig. \ref{fig:LinFFNN}. By choosing $\K$ as the output feedback law, the closed-loop dynamics of the LTI system \ref{eq:ltisys} becomes $\dot x = Ax + B\K(Cx) = (A+BK_N\cdots K_1C)x$,
and the LQR problem cost becomes
\begin{align}
    \label{eq:JOVPLQR}
    J(\K) 
    = \trace{P_\K \Sigma_0},
\end{align}
where for a given $\K$, $P_\K$ is the unique solution of the following Lyapunov equation:
\begin{align}
    \label{eq:lyapP}
    &P_\K(A+BK_N\dots K_1C)+(A+BK_N\dots K_1C)^\top P_\K\nonumber\\&+(K_N\dots K_1C)^\top R K_N\dots K_1C + Q = 0.
\end{align}
The notation $J(\K)$ and $J(K_1,K_2,\dots,K_N)$ are used interchangeably when the goal is to emphasize the dependency on the linear neural network $\K$ or on its parameters $(K_1,\dots,K_N)$. With this, consider the following problem definition.

\begin{definition}
    \label{def:Knnet}
    Let $\K$ be a LFFNN, and $A$, $B$, and $C$ be as in \eqref{eq:ltisys}. Define $\Kstbl:=\{\K~|~(A+BK_N\dots K_1C) \text{ is Hurwitz}\}$ and let $R\in\PD[m\times m]$ and $Q\in\PD[n\times n]$ be given symmetric positive definite matrices. Solving an overparameterized formulation of the model-free LQR problem consists in finding a $\Ks\in\Kstbl$ that solves
    \begin{mini*}|s|
        {\K\in\mathcal{K}}{J(\K):=\trace{P_\K\Sigma_0}}
        {}{}
        \addConstraint{ \eqref{eq:lyapP}.}
    \end{mini*}%
\end{definition}%
Then, a gradient flow for the overparameterized model-free LQR problem is defined for each $i=1,\dots,N$ and any fixed ``learning rate'' $\eta>0$ by imposing the following dynamics for the parameter matrices $K_i$ that compose $\K_0$
\begin{equation}
    \label{eq:GradFlowKi}
    \dot K_i = -\eta\frac{\partial J}{\partial K_i},
\end{equation}
and a candidate solution to the overparameterized model-free LQR problem is obtained by initializing the gradient flow at some $\K_0\in\Kstbl$ and selecting whichever point $\K_{ss}$ the solution converges to (assuming it converges to a point). It is evident that an equilibrium of the gradient flow dynamics \eqref{eq:GradFlowKi} is not necessarily the global optimum of the overparameterized poLQR, and a better understanding of the landscape of the problem is required before one can discuss the optimality of a solution obtained in such a manner. Nonetheless, $\dot K_i=0$ for all $i=1,\dots, N$ is a necessary condition for global optimality, which makes the equilibria of \eqref{eq:GradFlowKi} natural candidates for a optimal solution. Henceforth in this paper, it is assumed $\eta=1$, although comparisons between the proposed formulation and other formulations that explore variable values for $\eta$ could prove to be an interesting future direction of work.

Regarding the computation of the gradients of $J$ with respect to the matrices $K_i$, consider the following result:

\begin{lemma}
    \label{thm:jacCost}
    Let $B_i := BK_N\dots K_{i+1}$
    and $R_i := K_{i+1}^\top\dots K_N^\top RK_N\dots K_{i+1}$ for
    $i\in\{1,\dots,N-1\}$, $C_i := K_{i-1}\dots K_1C$ for
    $i\in\{2,\dots,N\}$, $B_N:=B$, $C_1:=C$, and $R_N:=R$. Then
    \begin{equation}
        \label{eq:gradientKi}
        \nabla_{K_i}J = 2[B_i^\top P_\K+R_iK_iC_i]L_\K C_i^\top,
    \end{equation}
    where $P_\K$ is the solution of \eqref{eq:lyapP}, $L_\K$ is the solution of
    \begin{align}
        &L_\K[A+BK_N\dots K_1C]^\top \nonumber\\&+ [A+BK_N\dots K_1C]L_\K+\Sigma_0 = 0,\label{eq:lyapL}
    \end{align}
    and $\Sigma_0$ relates to the distribution of initial conditions, being equal to the covariance matrix if the initialization is random Gaussian with zero mean, or equal to the identity for uniformly sampled unitary vectors.
\end{lemma}

Notice that we presented the results so far for arbitrary full-rank $C$ to keep the comparison with the results from \cite{rautert_computational_1997}, however moving forward we will assume full state feedback for the system, that is $C=I$, and initializations in the unit sphere, that is $\Sigma_0=I$. We next look at what can be said regarding convergence guarantees for the proposed problem.

\subsection{A conservation law for the overparameterized model-free LQR problem}
\label{sec:consvlaw}

Notice that, relative to the weight matrix of each hidden layer, the derivative of the cost $J$ relative to each parameter matrix, given by \eqref{eq:gradientKi} follows an iterative structure that allows the characterization of a conservation law that is satisfied by any solution. Such conservation law follows a very similar structure as the ones characterized for overparameterized linear regression (see for example Lemma 2.3 of \cite{chitour_geometric_2023}). This property is given in the following lemma:

\begin{lemma}
    \label{thm:invariance}
    For a gradient flow dynamics \eqref{eq:GradFlowKi} used for solving the overparameterized model-free LQR problem (poLQR) presented in Definition \ref{def:Knnet}, and for any $i$ from $1$ to $N-1$, the following quantity is invariant along any solution $(K_1(t),\dots,K_N(t))$ initialized in $\Kstbl$:
    \begin{align}
        \mathcal{C}_i:=& K_iK_i^\top - K_{i+1}^\top K_{i+1} \nonumber\\=& (K_iK_i^\top - K_{i+1}^\top K_{i+1})_{t=0}\label{eq:invexp},
    \end{align}
    where $\Ccal[i]$ are constant matrices of appropriate dimensions. We refer to the set $(\Ccal[1],\dots,\Ccal[N-1])$ as the set of invariants of a given solution.
\end{lemma}

A similar conservation law is leveraged to prove many of the properties of the overparameterized gradient flow for linear regressions, as can be seen from Lemma 2.3 in \cite{chitour_geometric_2023}, Lemma 1 in \cite{min_explicit_2021}, Lemma 2.1 of \cite{bah_learning_2022}, and others. The fact that such property also holds for the more general Linear Quadratic cost when overparameterized motivates the search presented in this paper for other useful properties that might hold for this case.

With this, and knowing that the LQR cost function is a rational function (see, for example, a discussion in \cite{sontag_remarks_2022}, section 4.3) the following result regarding the global convergence of solutions of \eqref{eq:GradFlowKi} can be stated:

\begin{theorem}
    \label{thm:existence}
    Any solution of the gradient flow \eqref{eq:GradFlowKi} initialized in $\Kstbl$ (defined as in Definition \ref{def:Knnet}): exists; is precompact; remains in $\Kstbl$ for all time; and converges to a critical point of the gradient flow dynamics.
\end{theorem}

This result not only guarantees invariance of the set of stabilizing neural networks and global convergence of solutions but also demonstrates how the invariance obtained in Lemma \ref{thm:invariance} can be used to extend results from the literature on overparameterized linear regressions to the context of the overparameterized model-free LQR problem. We next look at the case with $N=2$, \ie a single hidden layer, to enunciate an even stronger convergence result.

\subsection{Feedback control design with a single hidden layer}

Consider now the case where $N=2$ (single hidden layer). The literature on overparameterized linear regression is rich in results for this case, and this section aims to show that the main ones also hold for the design of optimal state feedback controllers.

We begin by proving that any critical point that is not a global minimum of the problem is necessarily a strict saddle. This result is, then, used to prove almost everywhere convergence to the global minimum of the problem. Then we characterize all critical points to discuss some intuition behind the problematic set of initializations, that is the set of initializations that do not converge to the global minimum. Let $\target$ be defined as in Section \ref{sec:ThrBknd}, then consider the following result: 

\begin{theorem}
    \label{thm:Saddles}
    Let $(K_1,K_2)$ be an equilibrium point of the gradient dynamics \eqref{eq:GradFlowKi}, then either

    \begin{itemize}
        \item The point $(K_1,K_2)$ is a global minimum of the system, \ie $K_2K_1\in\target$; or
        \item The point $(K_1,K_2)$ is a strict saddle of the dynamics, \ie the Hessian evaluated at $(K_1,K_2)$ has at least one negative eigenvalue.
    \end{itemize}
\end{theorem}

Because Theorem \ref{thm:Saddles} guarantees that the critical points are either strict saddles or global minima, and Theorem \ref{thm:existence} guarantees convergence to a critical point, we can apply Corollary \ref{cor:AEConvf} provided in Appendix \ref{apx:StrictSaddles} to get the following Corollary:

\begin{corollary}
    \label{cor:AEConv2Opt}
    For all initializations but a set of Lebesgue measure zero, the solution of the overparameterized gradient flow \eqref{eq:GradFlowKi} converges to a point $(K_1,K_2)$ such that $K_2K_1\in\target$, that is, almost all solutions initialized in $\Kstbl$ converge to an optimal feedback matrix and minimize \eqref{eq:JOVPLQR}.
\end{corollary}

Notice that Corollary \ref{cor:AEConv2Opt} is proven without needing to characterize the set of initializations that converge to a saddle. Such points are hard to characterize for an arbitrary saddle, although we can provide a characterization of the critical points themselves as follows:

\begin{lemma}
    \label{lem:critcharact}
    For the gradient flow \eqref{eq:GradFlowKi} with $N=2$ and $\kappa_1=\kappa>\max(m,n)$, and for any set of parameter matrices $(K_1,K_2)$ such that $K_2K_1\in\Kstbl$, the following are equivalent:
    
    \begin{enumerate}
        \item[1)] The point $(K_1,K_2)$ is  an equilibrium of \eqref{eq:GradFlowKi}, \ie$\dot K_1=K_2^\top 2[B^\top P_\K+RK_2K_1]L_\K=0,$ and $\dot K_2=2[B^\top P_\K+RK_2K_1]L_\K K_1^\top=0.$
        \item[2)] Let $\nabla_\K J := 2[B^\top P_\K+RK_2K_1]L_\K,$ then, there exist an SVD $\nabla_\K J(K_2K_1)=\Psi\Sigma\Phi^\top,$ and orthogonal matrices $\Gamma_{K_1}$,$\Gamma_{K_2}\in\mathbb{R}^{\kappa\times\kappa}$ such that: (a) $K_1=\Gamma_{K_1}\Sigma_{K_1}\Phi^\top$ and $K_2=\Psi\Sigma_{K_2}\Gamma_{K_2}^\top$ are SVDs of $K_1$ and $K_2$; and (b) $\Sigma\Sigma_{K_1}^\top=0$ and $\Sigma_{K_2}^\top\Sigma=0$.
    \end{enumerate}
\end{lemma}

From this Lemma, we can characterize the product $K_2K_1$ at critical points in terms of low-rank approximations of $K^*$ (the optimal LQR feedback matrix) as in the following corollary
\begin{corollary}
    \label{cor:finitesaddles}
    Let $K^*$ be the optimal value of $K\in\Kstbl$ that minimizes the LQR cost \eqref{eq:JLQR}. If $(K_1,K_2)$ is a critical point of the gradient flow dynamics \eqref{eq:GradFlowKi} with a $N=2$, then there exists an SVD of $K^*$ 
    \begin{equation*}
        K^* = \begin{bmatrix}
            \Psi_1^*, \Psi_2^*
        \end{bmatrix}\begin{bmatrix}
            \Sigma_1^* & 0 \\ 0 & \Sigma_2^*
        \end{bmatrix}\begin{bmatrix}
            (\Phi_1^*)^\top \\ (\Phi_2^*)^\top 
        \end{bmatrix},
    \end{equation*}
    with its singular values not necessarily in any order, such that
    \begin{equation*}
        K_2K_1 = \begin{bmatrix}
            \Psi_1^*, \Psi_2^*
        \end{bmatrix}\begin{bmatrix}
            \Sigma_1^* & 0 \\ 0 & 0
        \end{bmatrix}\begin{bmatrix}
            (\Phi_1^*)^\top \\ (\Phi_2^*)^\top 
        \end{bmatrix}.
    \end{equation*}
\end{corollary}

As an immediate consequence of Lemma \ref{lem:critcharact} and the consequent Corollary \ref{cor:finitesaddles}, one can notice that there is a finite number of values that the cost function \eqref{eq:JOVPLQR} can have at any critical point of the gradient flow dynamics. This characterizes a finite number of sets of critical points, as a function of the number of possible low-rank factorization of $K^*$. Despite that characterization, however, it is still hard to compute the center-stable manifold of the saddles, as we hope to illustrate next.

For some $p<\min(m,n)$, let $K^*_p$ denote a rank-$p$ factorization of $K^*$ and let the set of all $(K_1,K_2)$ such that $K_2K_1=K^*_p$ be given by $\target_p:=\{(K_1,K_2)~|~K_2K_1=K^*_p\}$. It is evident that for any $(K_1,K_2)\in\target_p$, $(K_1\mu,K_2(1/\mu))\in\target_p$ as well for any $\mu\neq 0$, and therefore $\target_p$ is continuous and unbounded. However, it is also easy to see that there exist two $(\bar K_1,\bar K_2)\in\target_p$ and $(\tilde K_1, \tilde K_2)\in\target_p$ for which there exist no $\mu\neq 0$ such that $(\bar K_1,\bar K_2)=(\mu\tilde K_1, (1/\mu)\tilde K_2)$.

The degrees of freedom for points in $\target_p$ come from the fact that multiple different values of $\Sigma_{K_1}$, $\Sigma_{K_2}$, $\Gamma_{K_1}$ and $\Gamma_{K_2}$ exist such that $K_2K_1=K_p^*$, however necessary and sufficient conditions on these matrices for the equality to hold do not exist to the authors' knowledge, which makes an analytic characterization of all points in a given set $\target_p$ difficult. This difficulty also explains why characterizing the center-stable manifold of the saddles is hard. Assume that for a given $(K_1,K_2)\in\target_p$ the center-stable manifold of that point is known, then to extend it to a ``neighborhood'' if the point in $\target_p$, one would need to be able to: first characterize all points arbitrarily close to $(K_1,K_2)$; and second derive how that characterization reflects in the characterization of the center-stable manifold of a point in $\target_p$. 

In this section, we have collected powerful results about the convergence of the gradient flow solution for the general problem and the single hidden layer case. These results provide some guarantee to the behavior of the solution but also illustrate some of the fundamental challenges of understanding deep and wide optimization formulations. We will follow up in the next section with a complete analysis of a simpler version of the problem, in the hopes of illustrating better some of the intuition derived from the results from this section.

\section{Analysis of the single-input/single-state case with one hidden-layer}
\label{sec:scalarcase}

To provide a better intuition behind the results given in the previous section, we now study a simple example of the considered problem. Assume $N=2$, $n=m=1$, but $\kappa_1=:\kappa$ arbitrary. The case where the parameters take these values is referred to as ``the vector case'', and if $\kappa=1$ then it is referred to as ``the scalar case''.

\begin{figure}[t]
    \centering
    {\includegraphics[scale=0.6]{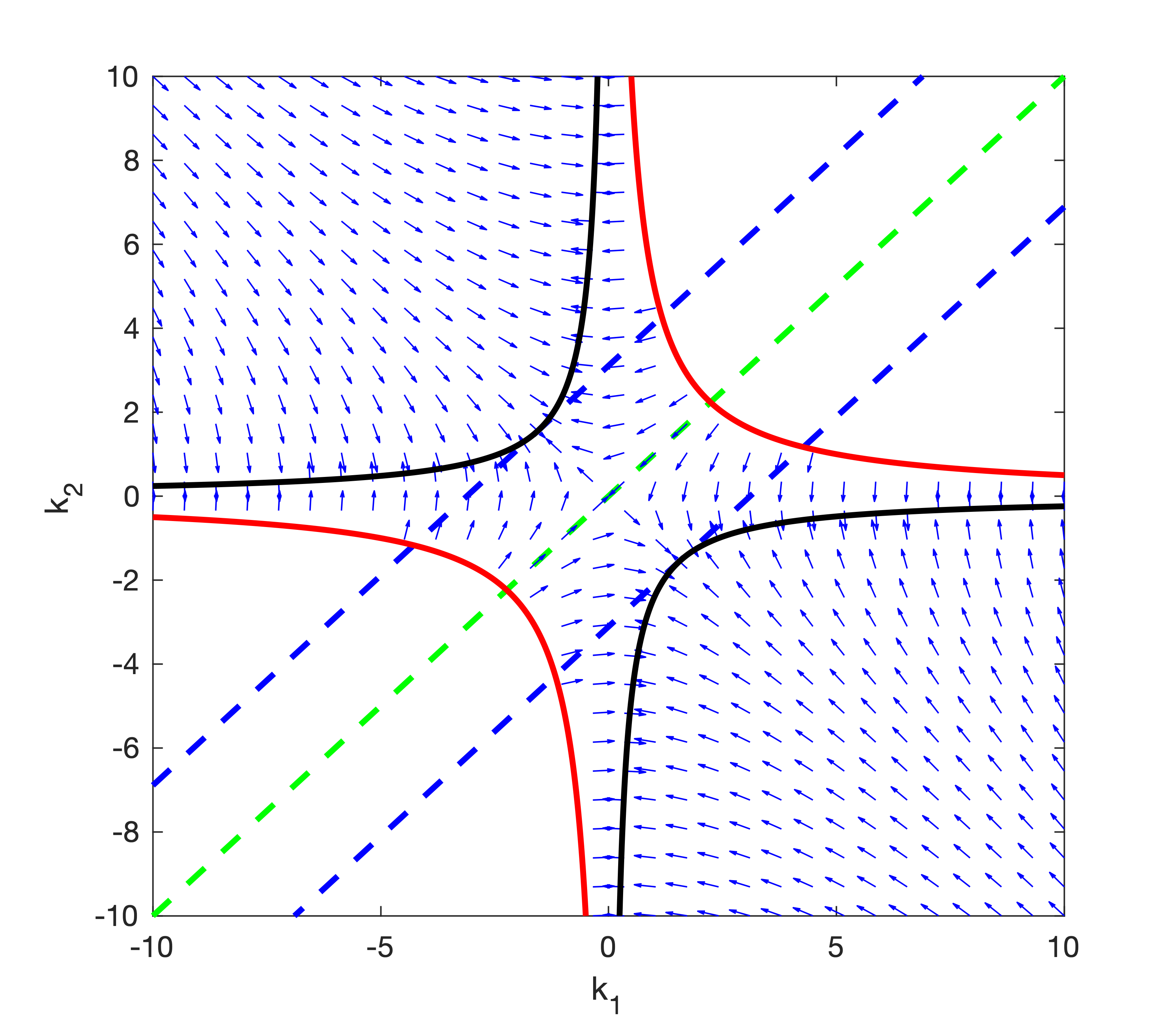}}
    \caption{\small Phase Plane for the gradient flow dynamics for the scalar case described in Section \ref{sec:scalarcase}, drawn for a stable $A$. The blue arrows depict the vector field at different points of the state space. The black hyperbolas are the new equilibria introduced by the condition $f(k_1,k_2)=0$, with $f(\cdot)$ as in \eqref{eq:k1dot} and \eqref{eq:k2dot}. The red hyperbolas are the borders of the set of $(k_1,k_2)$ such that $a+k_2k_1<0$, that is, such that the closed loop is stable. 
    The blue dashed lines are composed of the points that satisfy $d(k_1,k_2)=2\sqrt{|k_-^*|}$, while the green dashed line is the set for which $d(k_1,k_2)=0$ where $d(k_1,k_2)$ is as defined in Proposition \ref{prop:convergence}.}
    \label{fig:PPScalarCase}
\end{figure}

For the vector case, the system in consideration is of the form of \eqref{eq:ltisys} with $A,B\in\Reals$ and $x,u:\Reals_+\rightarrow\Reals$. Without loss of generality, assume $x(0)=1$, $B=1$, and denote $A=a$ to emphasize its scalar nature. Furthermore, assume the scalar weights for the cost \eqref{eq:JOVPLQR} are given by $Q=q>0$ and $R=r>0$, and the parameters to be optimized by $K_1=k_1\in\Reals[{\kappa\times 1}]$ and $K_2=k_2\in\Reals[{1\times \kappa}]$. Furthermore, the valid parameter space is defined as $\Kstbl:=\{(k_1,k_2)\in\Reals[{\kappa\times 1}]\times \Reals[{1\times\kappa}]~|~a+k_2k_1<0\}$. Assuming a feedback of the form $u=k_2k_1x$, with $(k_1,k_2)\in\Kstbl$ results in
\begin{align}
    J(k_1,k_2) &= \mathbb{E}_{x_0\in\mathcal{X}_0}\left[\int_0^\infty x(t)^2q+u(t)^2rdt\right] \nonumber\\
    &= \mathbb{E}_{x_0\in\mathcal{X}_0}\left[\int_0^\infty x(t)^2(q+(k_2k_1)^2r)dt\right] \nonumber\\
    &=\mathbb{E}_{x_0\in\mathcal{X}_0}\left[x(0)^2\right](q+(k_2k_1)^2r)\\&~~~~\times \int_0^\infty \mbox{e}^{2(a+k_2k_1)t}dt \nonumber\\ 
    &= -\frac{(q+(k_2k_1)^2r)}{2(a+k_2k_1)}.\label{eq:Jveccase}
\end{align}
Taking the gradient with respect to $k_1$ and $k_2$ gives
\begin{align}
    \nabla_{k_1}J(k_1,k_2) &= \label{eq:k1dot} f(k_1,k_2)k_2^\top \\ 
    \nabla_{k_2}J(k_1,k_2) &= f(k_1,k_2)k_1^\top ,\label{eq:k2dot}
\end{align}
where
\begin{equation*}
    f(k_1,k_2)=-\frac{r(k_2k_1)^2+2ark_2k_1-q}{2(a+k_2k_1)^2},
\end{equation*}
which, in turn, results in the following dynamics for the parameters
\begin{align}
    \dot k_1 &= -f(k_1,k_2)k_2^\top \label{eq:k1dotvec}\\
    \dot k_2 &= -f(k_1,k_2)k_1^\top.\label{eq:k2dotvec}
\end{align}
Notice that, similar to the observation made in \cite{de2023dynamics} for the vector case in linear regression, the vector dynamics of this problem is a simple nonlinear reparameterization of a linear dynamics. This means that inside $\Kstbl$, the phase plane should be that of a saddle with an inversion in the direction of the flow whenever $f<0$, and an extra equilibrium set given by $\{(k_1,k_2)\in\Kstbl~|~f(k_1,k_2)=0\}$. This can be observed graphically for the scalar case in the plot given by Fig. \ref{fig:PPScalarCase}.

The new equilibrium set given by $f(k_1,k_2)=0$ can be studied explicitly, this condition is satisfied for any $(k_1,k_2)\in\Kstbl$ that solves $r(k_2k_1)^2+2ark_2k_1-q=0$. The solutions to this quadratic equation are
\begin{align}
    k_2k_1 = -a+\sqrt{a^2+q/r} =: k^*_+ \label{eq:Zerosfp}\\
    k_2k_1 = -a-\sqrt{a^2+q/r} =: k^*_-\label{eq:Zerosfm},
\end{align}
with $k^*_+\not\in\mathcal{K}$ leaving $k^*_-$ as the only viable solution, which coincides with the optimal solution of the LQR problem for the scalar system, since from the theory on this problem one can write
\begin{align*}
    k^*_{\text{LQR}} = -R^{-1}B^\top P,
\end{align*}
where $P$ is the solution of
\begin{align*}
    A^\top P+PA-PBR^{-1}B^\top P+Q=0,
\end{align*}
which results in $K^*_{\text{LQR}}=k^*_-$ since $P>0$.

Furthermore, notice that $f(k_1,k_2)>0$ for all $(k_1,k_2)\in\mathcal{K}$ such that $k_2k_1>k^*_-$, since the positive root $k^*_+ = -a+\sqrt{a^2+q/r}>0$ is such that $a+k^*_+>0$, and the concavity of the parabola is negative. Also notice that if $k_2k_1<k^*$ then $f(k_1,k_2)<0$ by a similar argument.
However, notice that there is another equilibrium to this dynamics, given by $(k_1,k_2)=(0,0)$. For this equilibrium, $k_2k_1=0$ which is not the optimal solution of the LQR problem. Such equilibrium is referred to as a spurious equilibrium of the system and is only in $\mathcal{K}$ if $a<0$. Still, it is convenient to characterize a condition for which convergence to a global minimum is guaranteed. To do so, we adapt a result from \cite{de2023dynamics} to the design of feedback controllers:

\begin{proposition}
    \label{prop:convergence}
    For the overparameterized poLQR given by Definition \ref{def:Knnet} with $n=m=1$ and $N=2$ (\ie the vector case), the gradient flow solution converges to the global optimal value of the cost function \eqref{eq:JLQR} if and only if the gradient flow is initialized such that 
    \begin{equation*}
        d(k_1,k_2):=\|k_1-k_2^\top\|_2^2>0.
    \end{equation*}
\end{proposition}
Proposition \ref{prop:convergence} gives a necessary and sufficient condition for the convergence of a solution to the target set $\target:=\{(k_1,k_2)\in\Kstbl ~|~ k_2k_1=k^*\}$. For any point in $\target$, the value of the cost function $J(k_1,k_2)$ is the same, but that does not mean that all initializations that converge to $\target$ are equivalent. It was shown in Lemma \ref{thm:invariance} that different values for the conservation law are invariant along trajectories, so we show next how the values of this conservation law influence the convergence through the following definition and proposition..
\begin{definition}
    \label{def:vecimbalance}
    For the overparameterized poLQR given by Definition \ref{def:Knnet} with $n=m=1$ and $N=2$ (\ie the vector case), denote by $\mathcal{C}:=\mathcal{C}_1=k_1k_1^\top-k_2^\top k_2$, that is, the value of the invariant. Then, define the level of imbalance of a given solution as $c:=2\trace{\mathcal{C}^2}-\trace{\mathcal{C}}^2$.
\end{definition}
\begin{proposition}
    \label{prop:fastconv}
    For the overparameterized poLQR given by Definition \ref{def:Knnet} with $n=m=1$ and $N=2$ (\ie the vector case), let $(\phi_{k_1}(t,(k_{1},k_2)),\phi_{k_2}(t,(k_1,k_2)))$ be the solution to the gradient flow \eqref{eq:GradFlowKi} initialized at $(k_1,k_2)$ and let $\phi_J(t,(k_1,k_2)) = J(\phi_{k_1}(t,(k_{1},k_2)),\phi_{k_2}(t,(k_1,k_2)))$ be the trajectory of the cost function \eqref{eq:JOVPLQR} along a solution. For two distinct initializations $(\tilde k_1,\tilde k_2)$ and $(\bar k_1,\bar k_2)$ with levels of imbalance given by $\tilde c$ and $\bar c$ respectively, let
    \begin{itemize}
        \item $J(\tilde k_1, \tilde k_2)=J(\bar k_1, \bar k_2)$; and
        \item $|\tilde c|>|\bar c|\geq0$, with $\bar k_1\neq\bar k_2^\top$.
    \end{itemize}

    Then, for all time $t>0$ it follows that $\phi_J(t,(\tilde k_1,\tilde k_2))<\phi_J(t,(\bar k_1,\bar k_2))$. In other words, the cost converges faster to the minimum value for solutions initialized with a larger level of imbalance.
\end{proposition}
Proposition \ref{prop:fastconv} proves an increase in the rate of convergence for different solutions of the system, however, it provides no quantitative result, \ie it does not prove that the acceleration is unbounded. To further study advantages and trade-offs between different initializations, we next attempt to characterize the robustness of the solutions, \ie how the solutions can be expected to behave when the gradient is computed with an associated level of additive uncertainty.

Some intuition regarding the behavior of the solution under disturbance can be obtained from analyzing the scalar case. One can notice graphically from Fig. \ref{fig:PPScalarCase} that as $c$ increases, the associated equilibrium gets closer to the border of the set of stabilizing controllers, \ie the red and black hyperbolas in the figure ``meet at infinity''. At first sight, this can be a problematic observation when considering disturbances, as points in the target set can be arbitrarily close to the border of instability. However, this does not mean that any disturbance during the training can take the feedback matrix to instability. In fact, let $\delta\mathcal{K}$ be the border of $\mathcal{K}$ (\ie the red hyperbolas), and notice from \eqref{eq:k1dot} and \eqref{eq:k2dot} that in general, as $(k_1,k_2)\rightarrow\delta \mathcal{K}$, $|f(k_1,k_2)|\rightarrow \infty$, with its direction being away from the border. This means that only a disturbance of infinite magnitude on the training dynamics could take a solution initialized in $\mathcal{K}$ away from it.

To formalize this intuition, we prove the following ``ISS-type'' result regarding solutions of the overparameterized poLQR in the vector case when subject to additive uncertainties.
\begin{proposition}
\label{prop:ISS}
    For the overparameterized poLQR given by Definition \ref{def:Knnet} with $n=m=1$ and $N=2$ (\ie the vector case), consider solutions initialized in $\mathcal{K}$ and such that $\|k_1-k_2^\top\|_2\big |_{t=0}>2\sqrt{a_+}$, where $a_+=\max(0,a)$. Furthermore, let the dynamics be disturbed in the following form
    \begin{equation}
        \dot k_{1,2} = -\nabla_{k_{1,2}}J+u_{1,2},\label{eq:k1dotdist}
    \end{equation}
    where $u_1,u_2^\top:\mathbb{R}^+\rightarrow\mathbb{R}^\kappa$. Then for every $\epsilon>0$, there exists a $\delta>0$ such that if $\|u_1\|_\infty+\|u_2^\top\|_\infty\leq \delta$ then $\limsup_{t\rightarrow\infty}J(k_2(t)k_1(t))-J(k^*_-)\leq\epsilon$, where $\|\cdot\|_\infty$ is the infinity norm of a function.
\end{proposition}
Notice that the property characterized in Proposition \ref{prop:ISS} is not input-to-state stability as it is usually defined, and is more akin to a ``input-to-cost'' stability. Furthermore, due to the non-compactedness of the sets of critical points, one can even prove that for an arbitrarily small disturbance, the state will diverge, but will do so along a trajectory that will keep the value of the cost bounded. Nonetheless, in some sense this still guarantees that the solution remains ``close'' in the sense of the cost $J(\cdot)$ to the target set, even when subject to aditive disturbances.

Through this simple example, one can see how interesting and rich the problem discussed in this paper can be, as well as capture some of its intuition in a simpler context. The next section investigates numerically whether the increased speed of convergence, proven for the vector case here, might still hold for the general problem.

\section{Numerical results}
\label{sec:simul}

In this section, we investigate empirical distinctions between overparameterized and regular model-free LQR problems. The simulations were done using Matlab, and all code is available online in a repository \cite{paperRepo2025}. 
The selected $A$ and $B$ for the simulations are 
{\small\begin{align}
    A = -&\begin{bmatrix}
            5.2373 & 0.3452 & 0.6653 & 0.6715 & 0.3288 \\
            0.3452 & 5.4889 & 0.8060 & 0.3889 & 0.5584 \\
            0.6653 & 0.8060 & 5.0377 & 0.5735 & 0.5100 \\
            0.6715 & 0.3889 & 0.5735 & 5.3354 & 0.6667 \\
            0.3288 & 0.5584 & 0.5100 & 0.6667 & 5.4942
    \end{bmatrix}\\
    B^\top = &\begin{bmatrix}
        0 & 0 & 1 & 0 & 0 \\ 0 & 0 & 0 & 1 & 0 \\ 0 & 0 & 0 & 0 & 1
    \end{bmatrix}.
\end{align}}
All simulations are done for a 10-neuron single hidden layer neural network, since single hidden layer is enough to observe overparameterization and has better convergence guarantees. The choice of 10 hidden neurons was arbitrary.

For the gradient flow solution to be well-defined, a stabilizing initialization is required. Although this is a common necessary condition \cite{fazel2018global,mohammadi_convergence_2022}, as mentioned in the introduction, recent works in the literature \cite{zhang2023revisiting} explore the finite horizon formulation for the LQR problem to allow for arbitrary initializations. However, while studying the effects of overparameterization on such formulations could prove interesting, it is not in the scope of this paper.

Therefore, to generate the synthetic results that illustrate the distinct behaviors of an overparameterized formulation over the non-overparameterized formulation for the poLQR, we must first discuss the difference in the behavior of the solution based on the initialization.

\subsection{On the choice of Initialization}

Consider the phase plane of the scalar case depicted in Fig. \ref{fig:PPScalarCase} for reference. The first clear segmentation of the state space if the one done by the red hyperbolas, \ie between the values of $(K_1,K_2)$ such that $A+BK$ is Hurwitz or not.

Another similar segmentation is done by the black hyperbolas in the same figure Notice that any solution initialized in between the two hyperbolas will never cross either hyperbola, and vice-versa. This happens because at the black hyperbolas both gradients $\nabla_{K_1}J(K_1,K_2)=0$ and $\nabla_{K_2}J(K_1,K_2)=0$, and by continuity of the solution, it cannot cross over.

Finally, the quadrants also separate the state space in two, where any initialization in the second and fourth quadrants always converges to the global optimum (black hyperbola), while initializations in the first and third quadrants can converge to the saddle at the origin.

\begin{figure}
    \centering
    \begin{tikzpicture}[scale=1]
        \begin{axis}[
            xmin = -3, xmax = 3,
            ymin = -3, ymax = 3,
            zmin = 0, zmax = 1,
            axis equal image,
            view = {0}{90},
            xtick = \empty,
            ytick = \empty,
            xlabel={\footnotesize $K_1$},
            ylabel={\footnotesize $K_2$},
            no markers,
            every axis plot/.append style={ultra thick}
        ]
            
            \fill [yellow!10!white, domain=-3:-1/3, variable=\x]
              (-3, 0)
              -- plot ({\x}, {1/\x})
              -- (0, -3)
              -- (0, 0)
              -- cycle;
              \fill [yellow!10!white, domain=1/3:3, variable=\x]
              (0, 0)
              -- (0, 3)
              -- plot ({\x}, {1/\x})
              -- (3,0)
              -- (0, 0)
              -- cycle;

              \fill [green!10!white, domain=-3:-1/3, variable=\x]
              (-3, 0)
              -- plot ({\x}, {-1/\x})
              -- (0, 3)
              -- (0, 0)
              -- cycle;
              \fill [green!10!white, domain=1/3:3, variable=\x]
              (0, 0)
              -- (0, -3)
              -- plot ({\x}, {-1/\x})
              -- (3,0)
              -- (0, 0)
              -- cycle;

              \fill [blue!10!white, domain=-3:-1/3, variable=\x]
              (-3, 3)
              -- plot ({\x}, {-1/\x})
              -- cycle;
              \fill [blue!10!white, domain=1/3:3, variable=\x]
              (3, -3)
              -- plot ({\x}, {-1/\x})
              -- cycle;

              \fill [red!10!white, domain=-3:-1/3, variable=\x]
              (-3, -3)
              -- plot ({\x}, {1/\x})
              -- cycle;
              \fill [red!10!white, domain=1/3:3, variable=\x]
              (3, 3)
              -- plot ({\x}, {1/\x})
              -- cycle;
            
            \addplot[red!90!black,domain=-3:-1/3] {1/(x)};
            \addplot[red!90!black,domain=1/3:3] {1/(x)};
            \addplot[black,domain=-3:-1/3] {-1/(x)};
            \addplot[black,domain=1/3:3] {-1/(x)};
            \draw[ultra thick] (0,-3) -- (0,3);
            \draw[ultra thick] (-3,0) -- (3,0);

            \draw[thick,fill=white] (-2,2) circle [radius=0.2] node {\tiny I};
            \draw[thick,fill=white] (2,-2) circle [radius=0.2] node {\tiny I};
            \draw[thick,fill=white] (-1/2,1/2) circle [radius=0.2] node {\tiny II};
            \draw[thick,fill=white] (1/2,-1/2) circle [radius=0.2] node {\tiny II};
            \draw[thick,fill=white] (-2,-2) circle [radius=0.2] node {\tiny IV};
            \draw[thick,fill=white] (2,2) circle [radius=0.2] node {\tiny IV};
            \draw[thick,fill=white] (-1/2,-1/2) circle [radius=0.2] node {\tiny III};
            \draw[thick,fill=white] (1/2,1/2) circle [radius=0.2] node {\tiny III};
        \end{axis}
    \end{tikzpicture}
    \caption{\small Depiction of the four different regions of the state-space based on the expected behavior of the solution. 
    }
    \label{fig:reghionsstatespace}
\end{figure}

From this informal analysis, one can draw Fig. \ref{fig:reghionsstatespace}, which can be expected to describe the behavior of the solution to some degree, even if not extensively. We will perform the simulations for initializations $(K_1(0),K_2(0))=(K_{10},K_{20})$ such that $K_{20}K_{10}=\eta K^*$ where $K^*$ is the optimal feedback matrix and $\eta$ is a scalar. Intuitively, one would expect that if $\eta>1$, then the system would be initialized in a region of the state-space analogous to \circled{I} in Fig. \ref{fig:reghionsstatespace}, where $\eta$ can be arbitrarily large and the solution still converges to the target set. Similarly, if $1>\eta>0$, the solution is in a region analogous to \circled{II}, and the closer $\eta$ is to $0$, the longer the initialization should take to converge to the target set. Finally, if $\eta<0$ then it is in a region analogous to \circled{III}, or if $|\eta|$ is too large, then the solution does not exist.

To be more specific, for any given desired $\eta$ we compute $K^*$ first, then compute a SVD for it as $K^*=\Psi\Sigma\Phi$ and a random orthogonal $10\times 10$ matrix $\Gamma$. Then, we define $K_{10}=\sign(\eta)\sqrt{|\eta|}\mu\Gamma\Sigma^{1/2}\Phi^\top$ and $K_{20} = (\sqrt{|\eta|}/\mu)\Psi\Sigma^{1/2}\Gamma^\top$ for $\mu$ varying from $1$ to $100$ defining more or less imbalanced initializations for the same $\eta$.

As mentioned before, this does not encompass all possible behaviors for the solution of the general case with a single hidden layer. To illustrate this fact, we will perform simulations for all three cases described above ($\eta>1$, $1>\eta>0$ and $\eta<0$) and a final simulation for an initialization selected specifically to not lie in any of the regions described by the different values of $\eta$.

After an overview of the behavior of the solution is provided, we will investigate how overparameterization affects the convergence in a scenario where the gradient is numerically estimated from evaluations of the cost function, resulting in imprecise approximations and introducing uncertainty to the dynamics.

\subsection{Results with the exact gradient}

\begin{figure}[t]
    \centering
    \includegraphics[trim={0cm 0 0 0},width=1\linewidth]{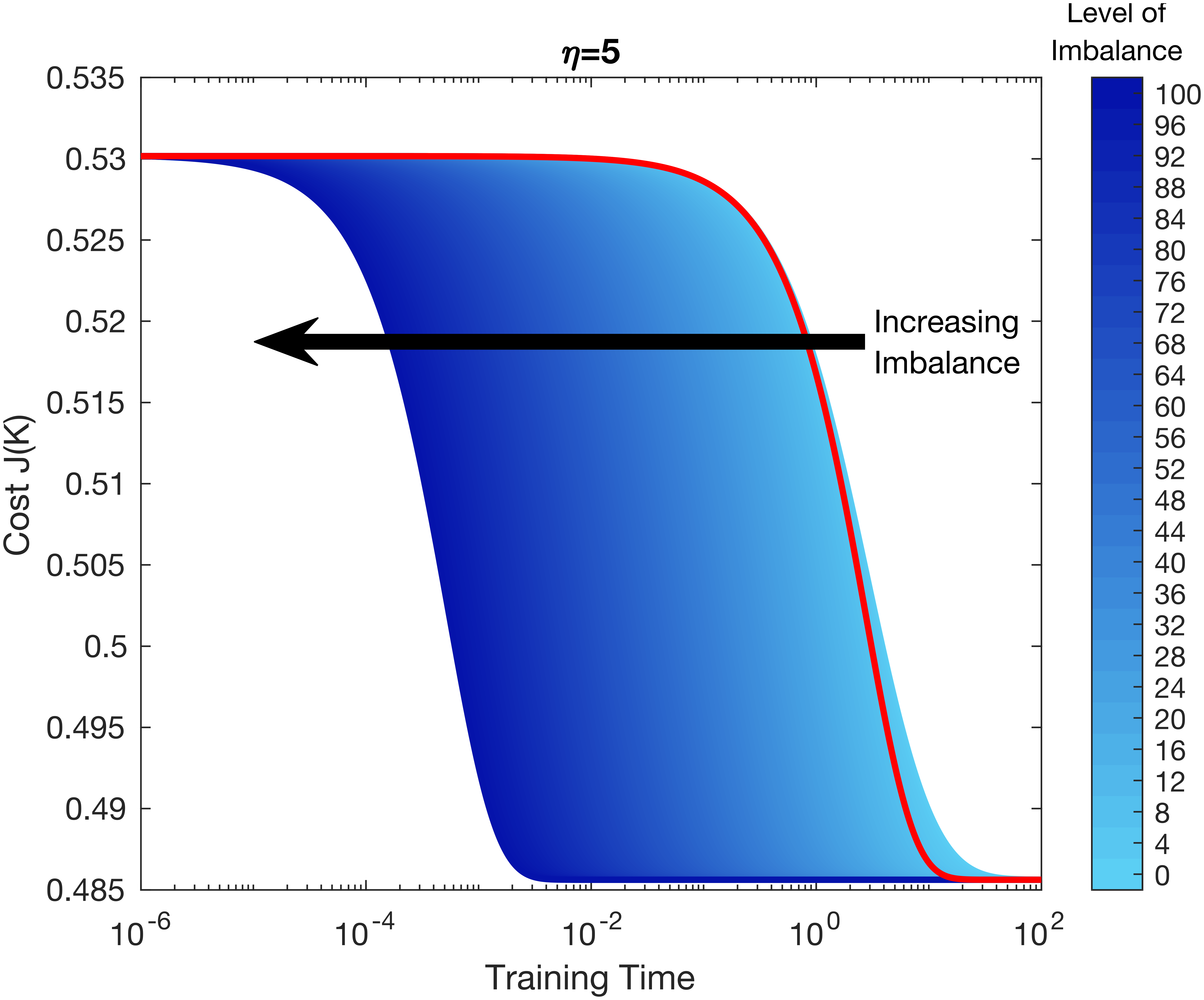} (a) \\
    \includegraphics[trim={0cm 0 0cm 0},width=1\linewidth]{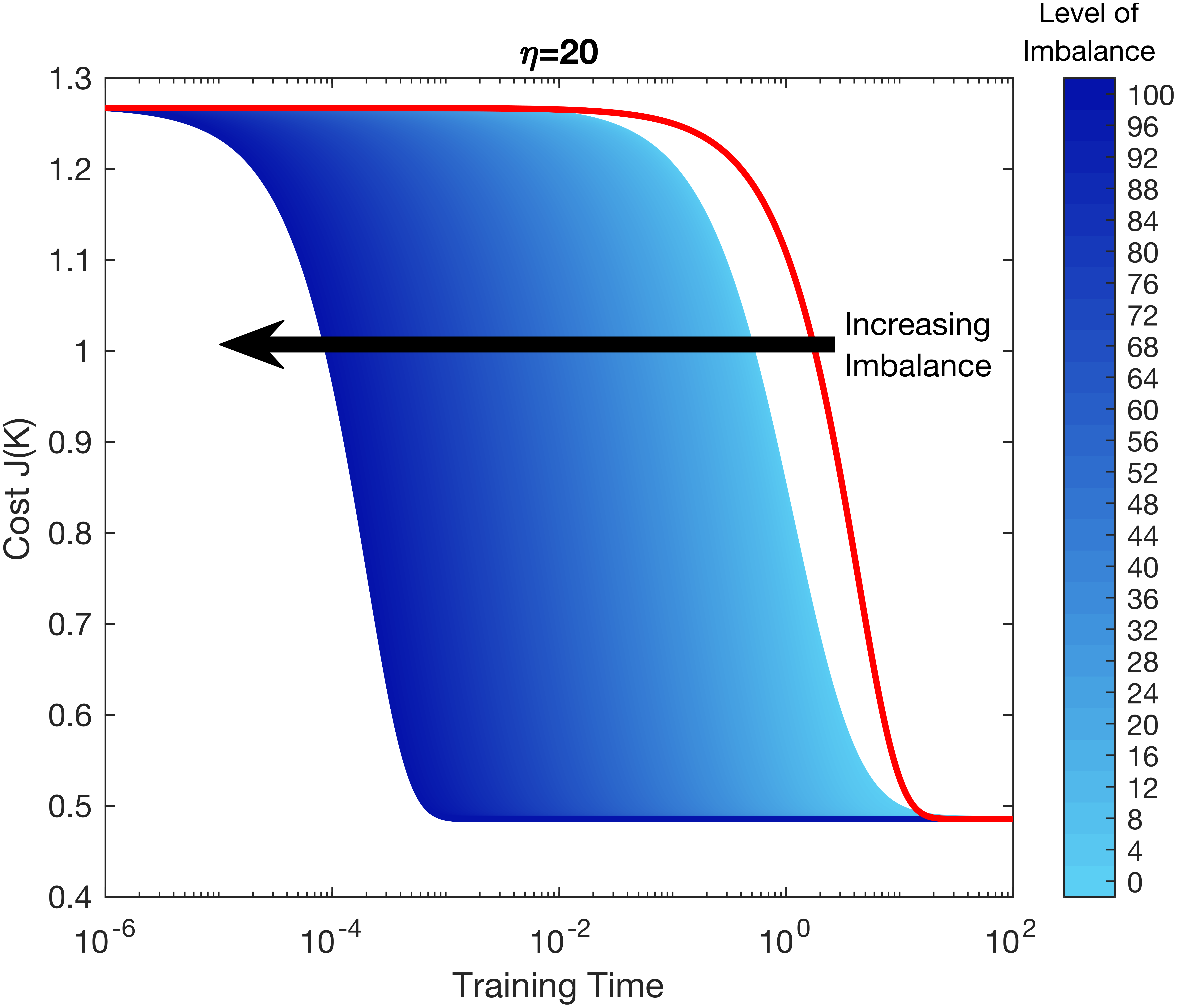} (b)
    \caption{Simulations done for initializations with $\eta>1$. Solutions were initialized with $\eta=5$ in (a) and with $\eta=20$ in (b). Solutions from light to dark blue depict overparameterized solutions with different levels of imbalance $\mu\in[1,100]$, and the red curve shows the non-overparameterized solution. 
    }
    \label{fig:SimulEtaL1}
\end{figure}

\begin{figure}[t]
    \centering
    \includegraphics[trim={0 0 0 0},width=1\linewidth]{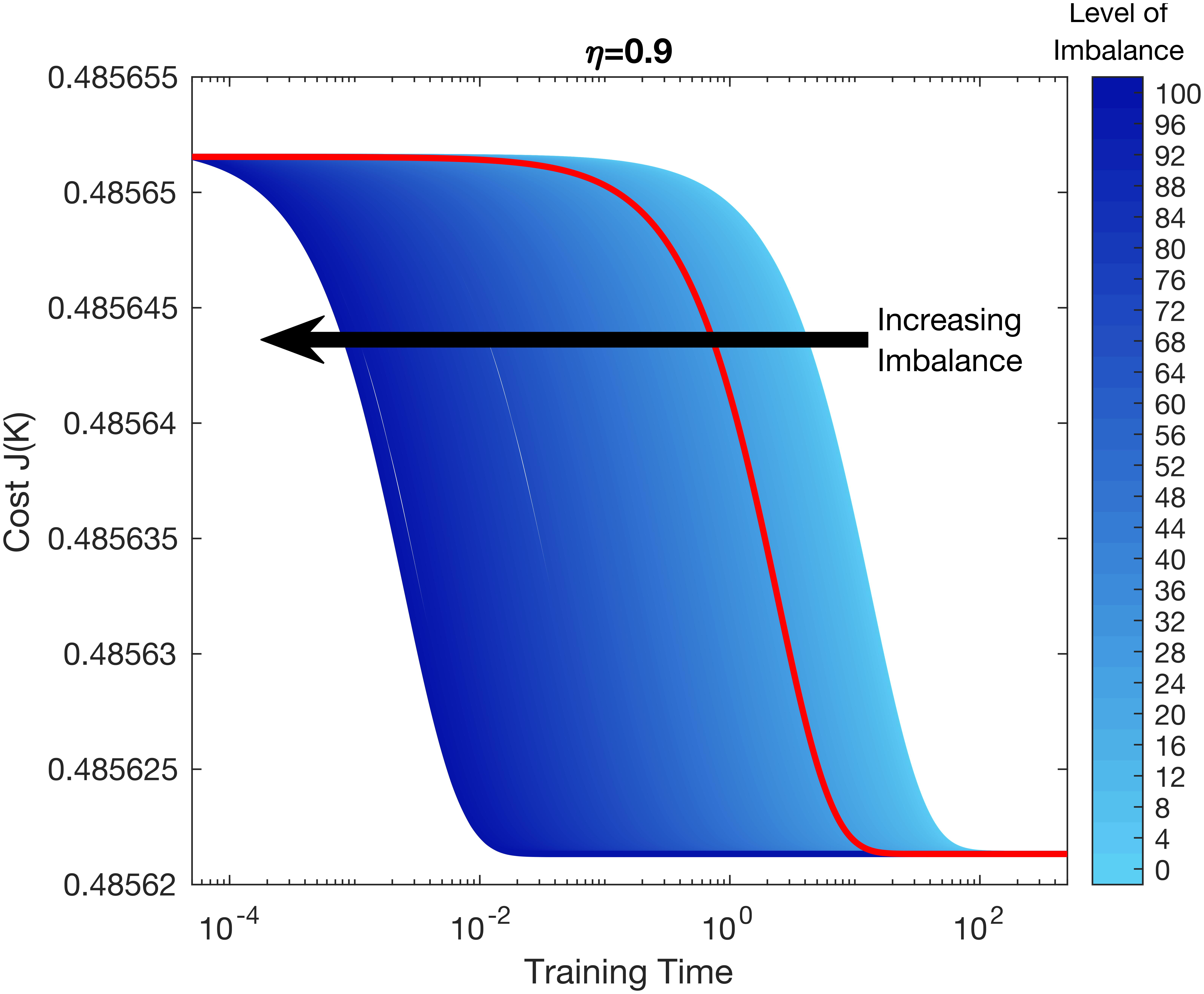} (a) \\
    \includegraphics[trim={0cm 0 0cm 0},width=1\linewidth]{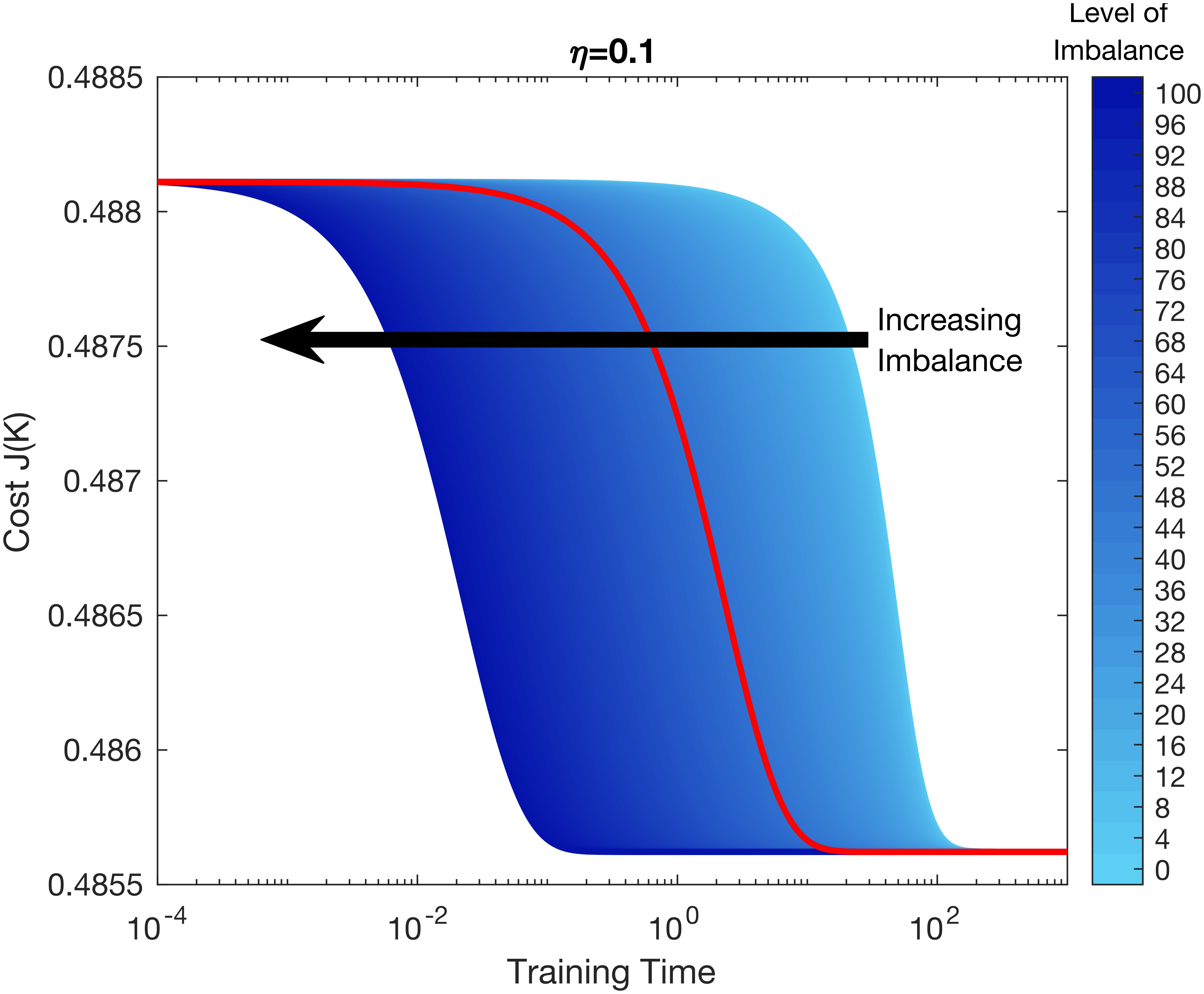} (b)
    \caption{Simulations done for initializations with $1>\eta>0$. Solutions were initialized with $\eta=0.9$ in (a) and $\eta=0.1$ in (b). Solutions from light to dark blue depict overparameterized solutions with different levels of imbalance $\mu\in[1,100]$, and the red curve shows the non-overparameterized solution. 
    }
    \label{fig:Simul1LEtaL0}
\end{figure}

\begin{figure}[t]
    \centering
    \includegraphics[trim={0cm 0 0 0cm},width=1\linewidth]{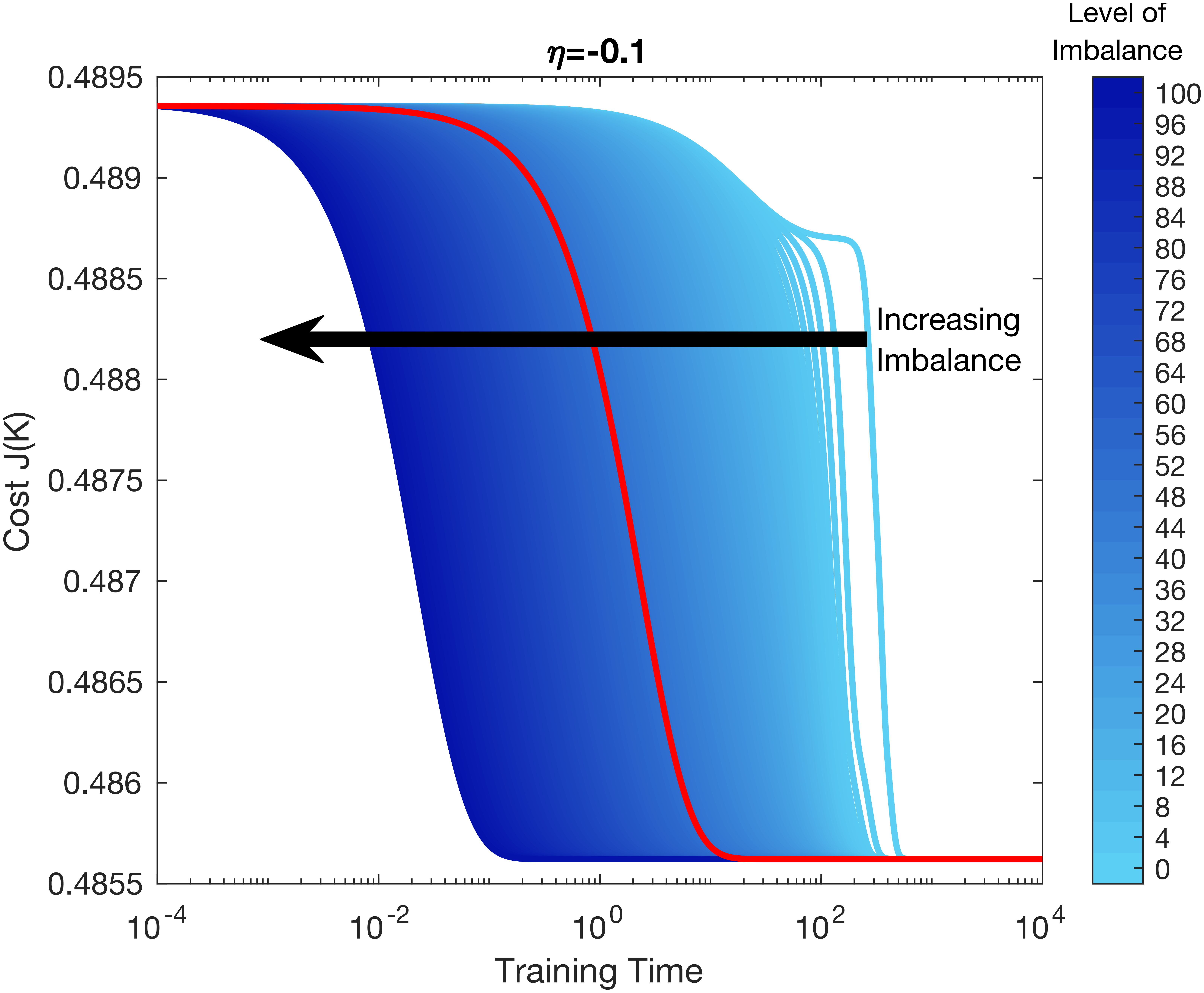} (a)\\
    \includegraphics[trim={0cm 0 0cm 0},width=1\linewidth]{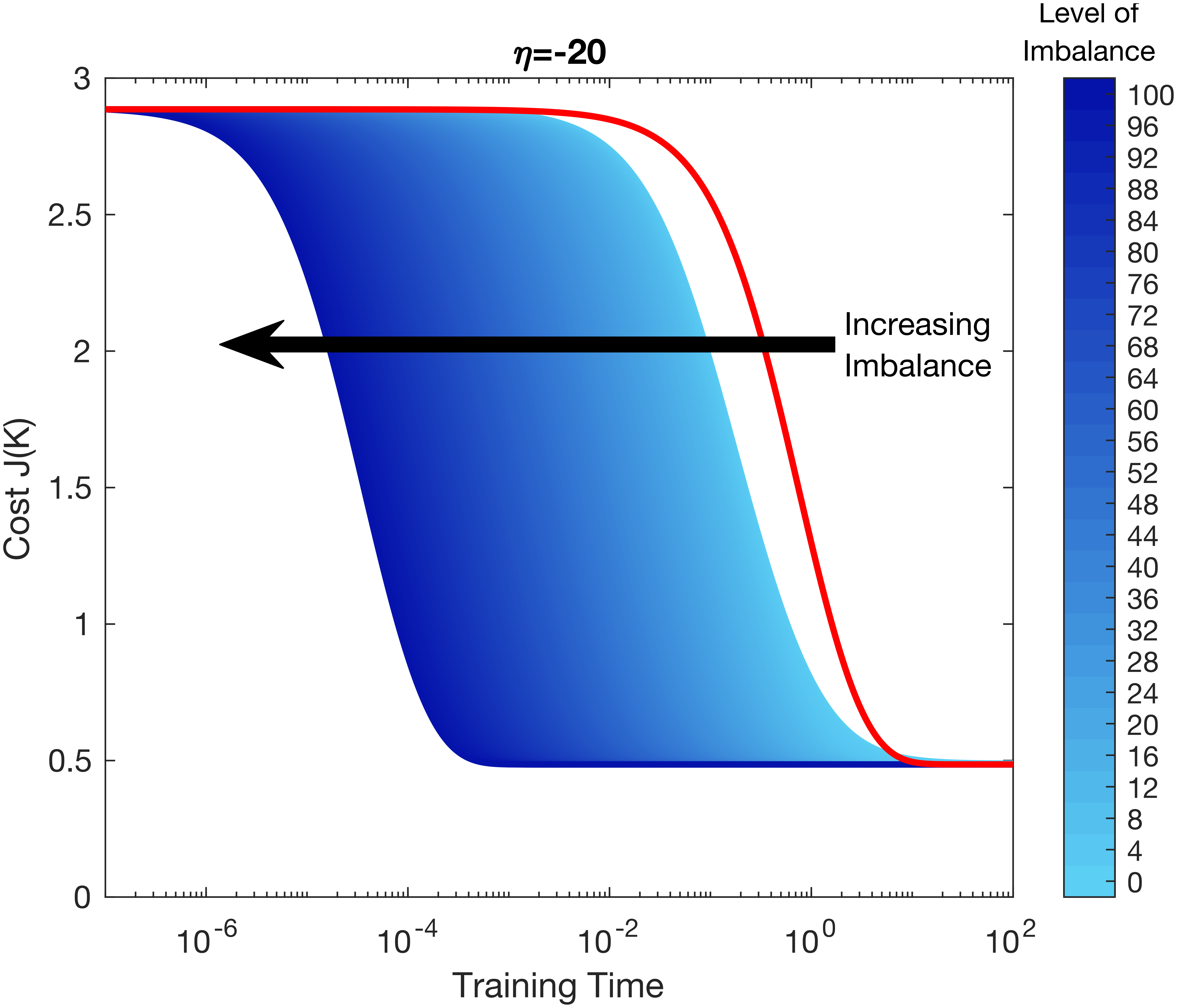} (b)
    \caption{Simulations done for initializations with $\eta<0$. Solutions were initialized with $\eta=-0.1$ in (a) and with $\eta=-20$ in (b). Solutions from light to dark blue depict overparameterized solutions with different levels of imbalance $\mu\in[1,100]$, and the red curve shows the non-overparameterized solution. 
    }
    \label{fig:SimulEtaL0}
\end{figure}

In this section, we will discuss the numerical simulation results for the gradient flow for the case when the gradient is perfectly known. The simulations are done for two different initializations with $\eta>1$ (Fig. \ref{fig:SimulEtaL1}), two with $1>\eta>0$ (Fig. \ref{fig:Simul1LEtaL0}), and two with $\eta<0$ (Fig. \ref{fig:SimulEtaL0}). Finally, two other simulations are done with a different initialization to illustrate a distinct behavior of the solution (Fig. \ref{fig:SimulEtaweird_zoom}).

For the simulations with $\eta>1$ in Fig. 6, notice how the overparameterized solutions (shades of blue) converge relative to the non-overparameterized solution (red) depending on how far from the optimum the solution is initialized. For $\eta=5$, the slowest of the overparameterized solutions (lightest blue) converges almost as quickly as the non-overparameterized solution but is overtaken as the solutions get closer to the optimum. Nonetheless, with an arguably small value for the imbalance term $\mu$, it is verifiable that an overparameterized solution will converge more quickly than the non-overparameterized one. This becomes even more evident for the solutions initialized with $\eta=20$, where all overparameterized solutions converge to the optimum faster than the non-overparameterized solution. Furthermore, notice how the solution to the overparameterized gradient flow has a different profile than the non-overparameterized solution, indicating that the overparameterized formulation did not simply accelerate the convergence, but changed the behavior of the solution.

For the simulations with $1>\eta>0$ in Fig. \ref{fig:Simul1LEtaL0}, the variation in the values of the cost function is limited by the values of $J(K)$ for $K=K^*$ and $K=0$. Furthermore, notice that the solutions initialized with $\eta=0.9$ converge generally faster than the ones initialized with $\eta=0.1$. This happens because as $\eta\rightarrow 0$, the initialization approaches the saddle, slowing down. Despite that, however, a big enough imbalance can always be imposed to generate a solution that converges more rapidly than the non-overparameterized solution.

Next, for the simulations with $\eta<0$ depicted in Fig. \ref{fig:SimulEtaL0}, before we can discuss the simulation results we first need to argue that theoretically for any initialization with $\eta<0$ and in $\Kstbl$, if $\mu=1$ then the resulting solution should converge to the saddle-point at the origin. To show this, first notice that for $\mu=1$, $\mathcal{C}=0$ by construction of the initialization. Then, notice that if $\eta<0$, then $J(K_{20}K_{10})>J(0)$. This can be shown theoretically, but for the simplicity of this analysis, this was verified numerically for this specific example. Next, since $J(K_{20}K_{10})>J(0)>J(K^*)$, by continuity any solution initialized at $(K_{10},K_{20})$ must pass through a point such that $K_{20}K_{10}=0$ before it can reach the target set $\target$. Finally notice that the only point such that $\mathcal{C}=0$ and that $K_2K_1=0$ is the origin, which is a saddle of the dynamics. 

\begin{figure}[t]
    \centering
    \includegraphics[trim={0cm 0 0cm 0},width=1.1\linewidth]{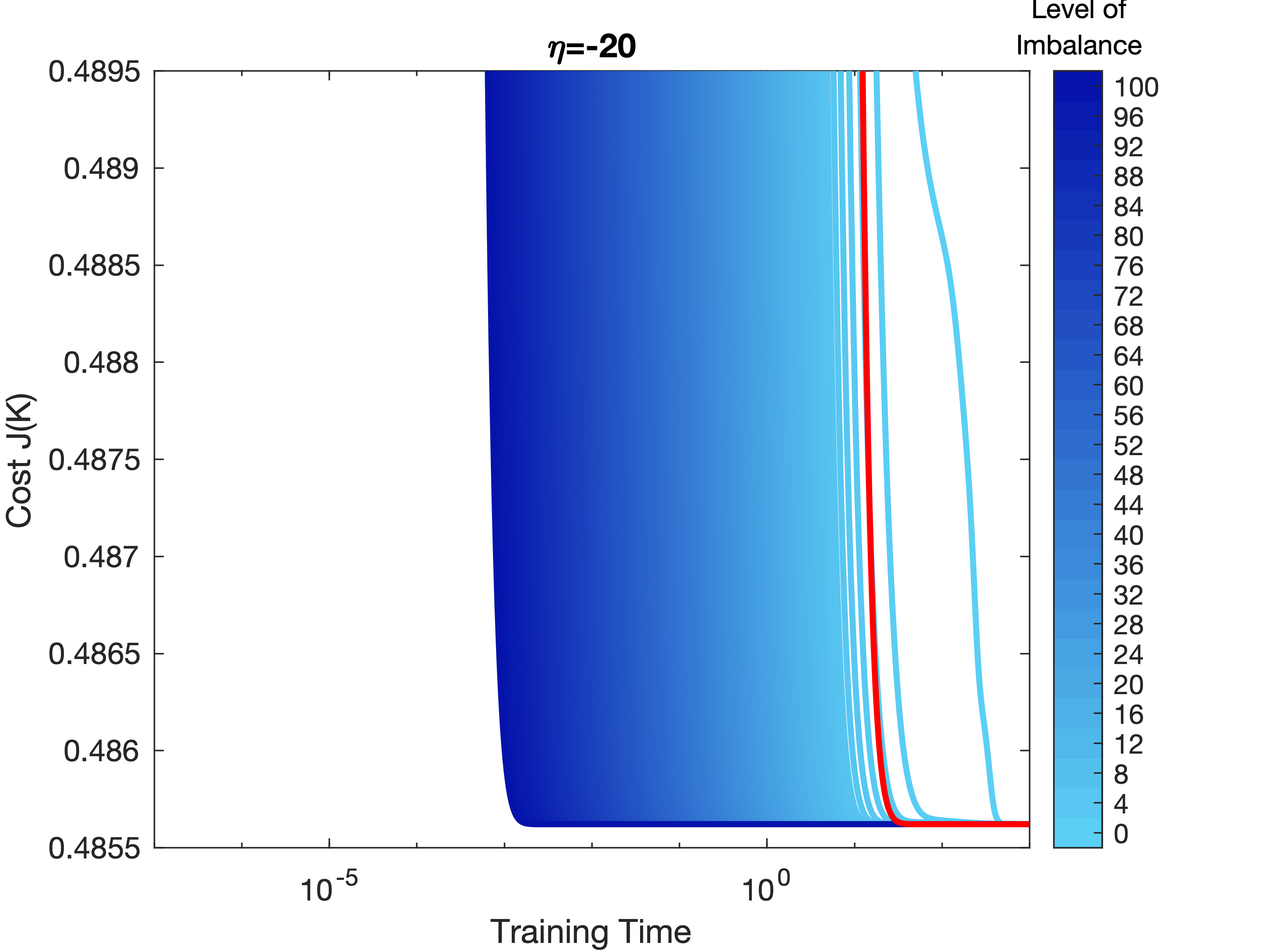} (a)\\
    \includegraphics[trim={0cm 0 0cm 0},width=1\linewidth]{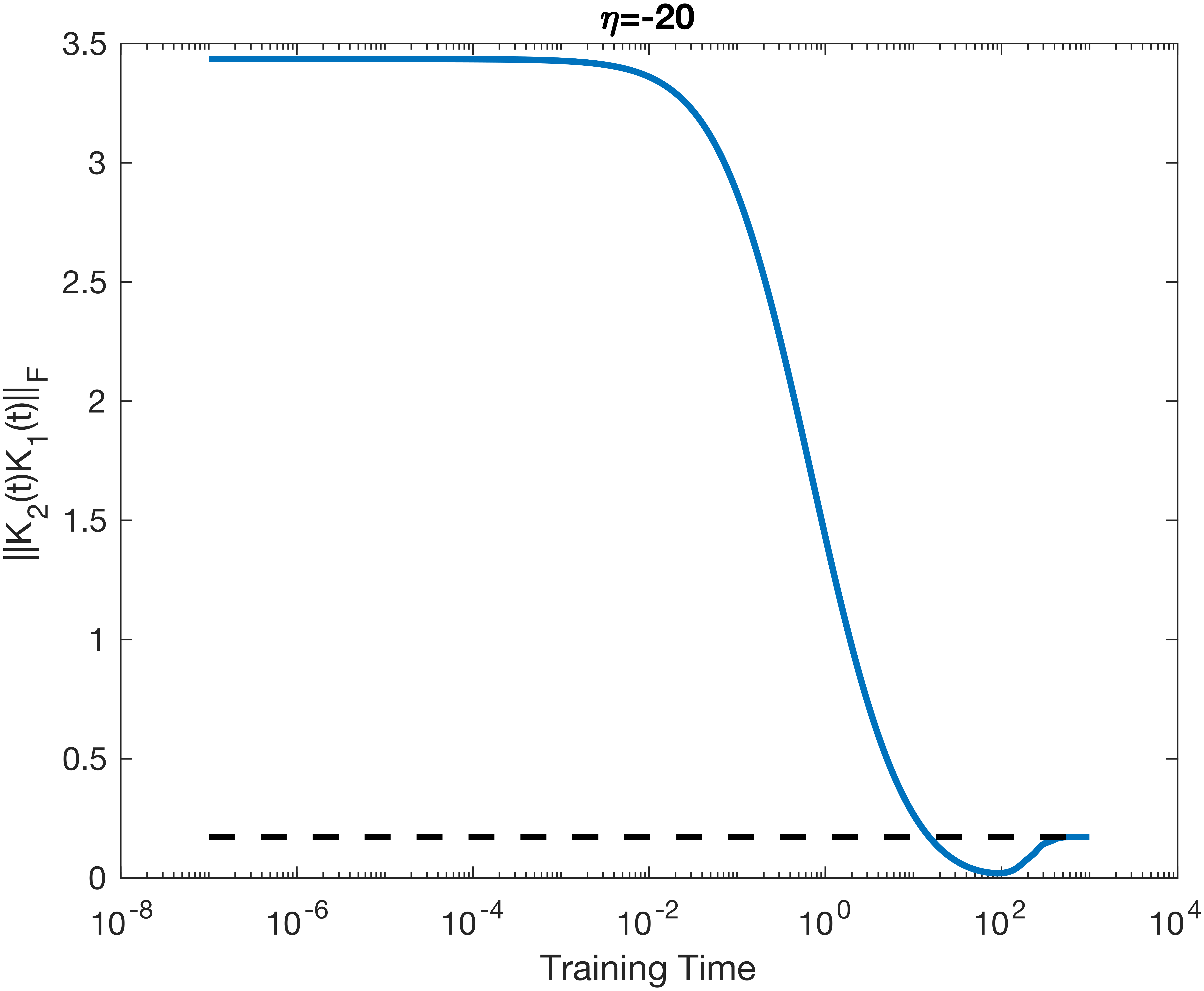} (b)
    
    \caption{Simulations done for initializations with $\eta=-20$. In (a) we have a zoomed version of the right graph in Fig. \ref{fig:SimulEtaL0}, where the influence of the saddle in the imbalanced solution becomes more evident. In (b) we have a plot of the Frobenius norm of the product $K_2(t)K_1(t)$, showing that the solution comes very close to $K_2K_1=0$, but then converges to the dashed line, which is the Frobenius norm of $K^*$.}
    \label{fig:SimulEtaL0_zoom}
\end{figure}

\begin{figure}[t]
    \centering
    \includegraphics[trim={0cm 0 0 0},width=1\linewidth]{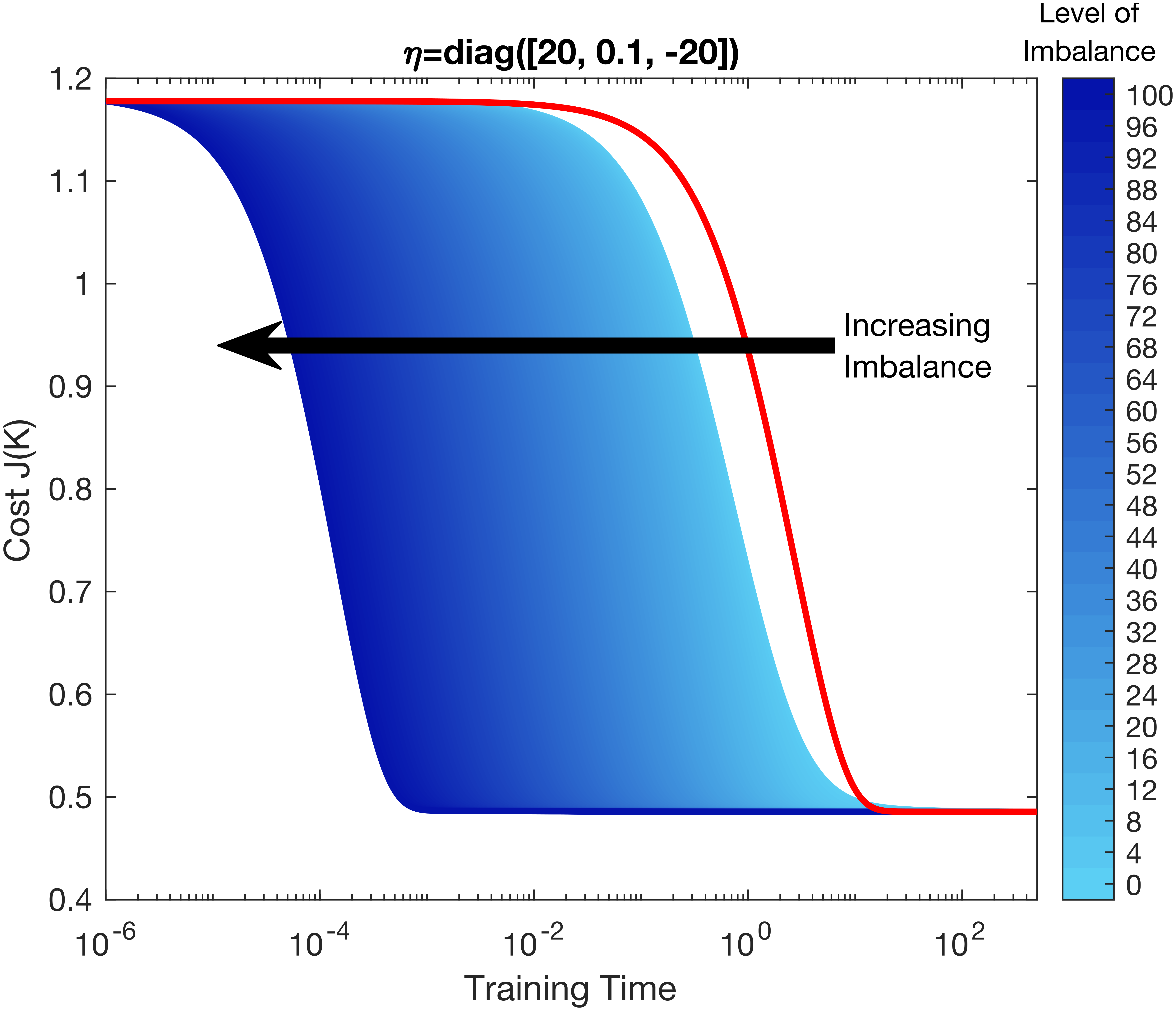} (a) \\
    \includegraphics[trim={0cm 0 0cm 0},width=1.09\linewidth]{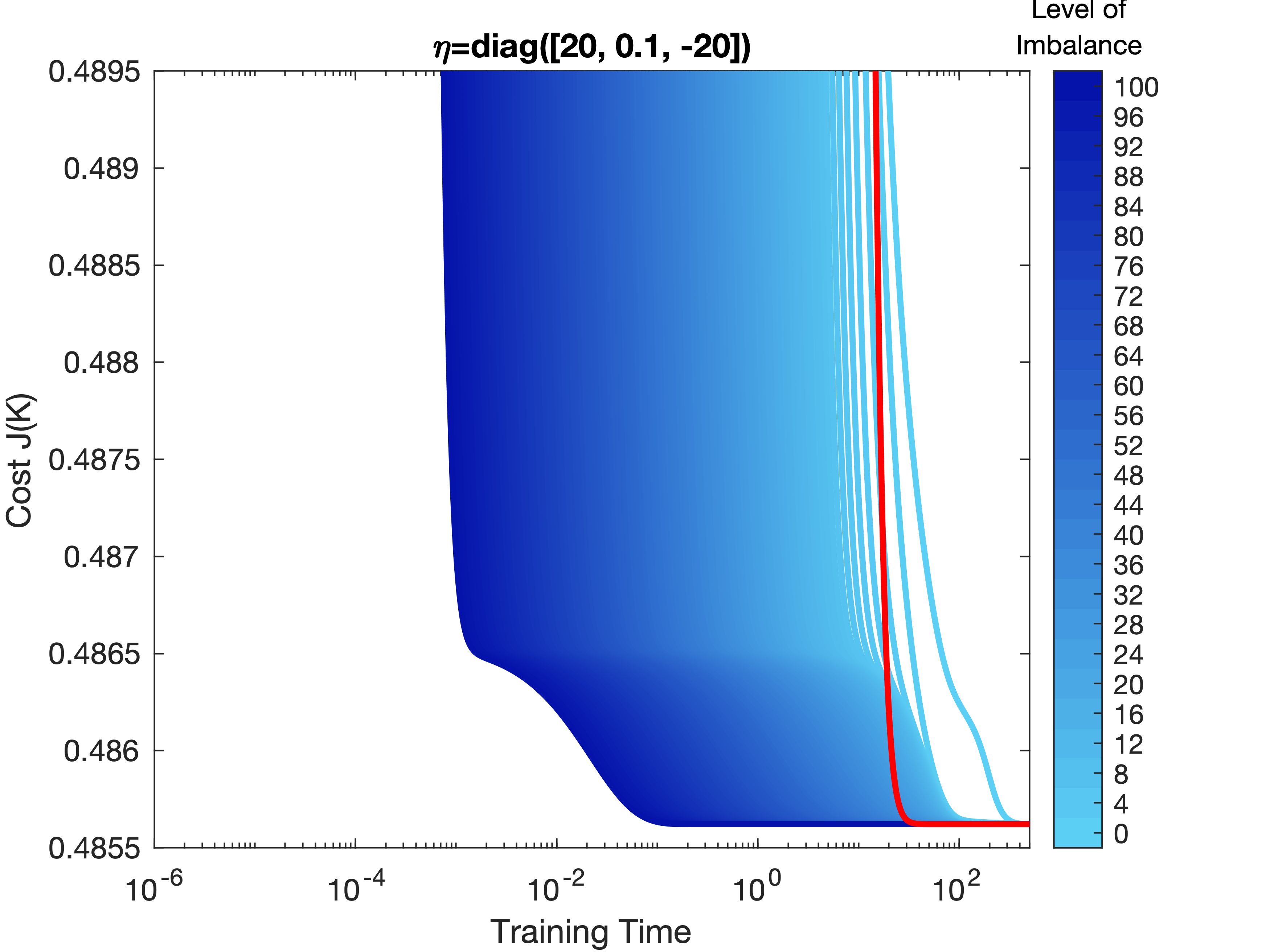} (b)
    
    \caption{Simulations done for initializations with $\eta=\mbox{diag}([20,0.1,-20])$. On (a) we have the entire trajectory for the solutions and on (b) we have a zoomed version of the plot. Notice that despite this initialization not lying in any of the pre-identified regions of the state space, many of the qualitative observations we made for the behavior of the solution still hold.}
    \label{fig:SimulEtaweird_zoom}
\end{figure}

This explains the strange behavior of the solutions initialized with $\mu=1$ and $\eta=-0.1$ in Fig. \ref{fig:SimulEtaL0}, where the solution looks like it is converging to a suboptimal value for the cost function. However, despite theoretically converging to the saddle at $0$, the simulation solution eventually escapes it due to accumulated errors in the numerical simulation, and reaches the global minimum. When looking at solutions initialized at $\eta=-20$ one might think a priory that the same phenomenon observed when $\eta=-0.1$ does not happen, however, if one looks at the zoomed graph in Fig. \ref{fig:SimulEtaL0_zoom}(a), one can see clearly that the solution initialized with $\eta=-20$ and $\mu=1$ is affected by the proximity to the saddle, although less than when initialized with $\eta=-0.1$. This effect is even more evident if we look at Fig. \ref{fig:SimulEtaL0_zoom}(b), which depicts the Frobenius norm of $K_2(t)K_1(t)$ along the solution with $\mu=1$. Notice that the norm of the matrix product approaches zero, but eventually escapes the saddle due to accumulated numerical errors.

Finally, we present a set of simulations selected specifically to not fit in any of the previously discussed cases. To do that, notice that $K^*$ has three singular values, so instead of multiplying all three by the same $\eta$, we multiply the first one by $20$, the second by $0.1$, and the third by $-20$. The resulting solutions are shown in Fig. \ref{fig:SimulEtaweird_zoom}. Notice that despite this initialization not lying in any of the pre-identified regions of the state space, many of the qualitative observations made for the behavior of the solution still hold. Furthermore, the saddle that the solutions approach in this case is not the origin (which is an isolated critical point), but a non-compact set of saddles, which explains why the effect of the proximity to the saddle affects all solutions, regardless of the level of imbalance.

We conclude this section of simulations with exact knowledge of the value of the gradient with a final observation regarding the level of imbalance. Theoretically, there is no limit to how imbalanced one can make an initialization, however, in practice, the more imbalanced an initialization, the stiffer the resulting ODE, making it harder for numerical solvers for ordinary differential equations to simulate the system. Therefore, although the gradient flow converges ``more quickly'' in simulation time, the stiff ODE starts to take longer to solve in practice if the initialization is chosen to be too imbalanced. This poses a real-life trade-off on how imbalanced one can make the initialization.

\subsection{Results with uncertain gradient}

\begin{figure}
    \centering
    \includegraphics[width=1\linewidth]{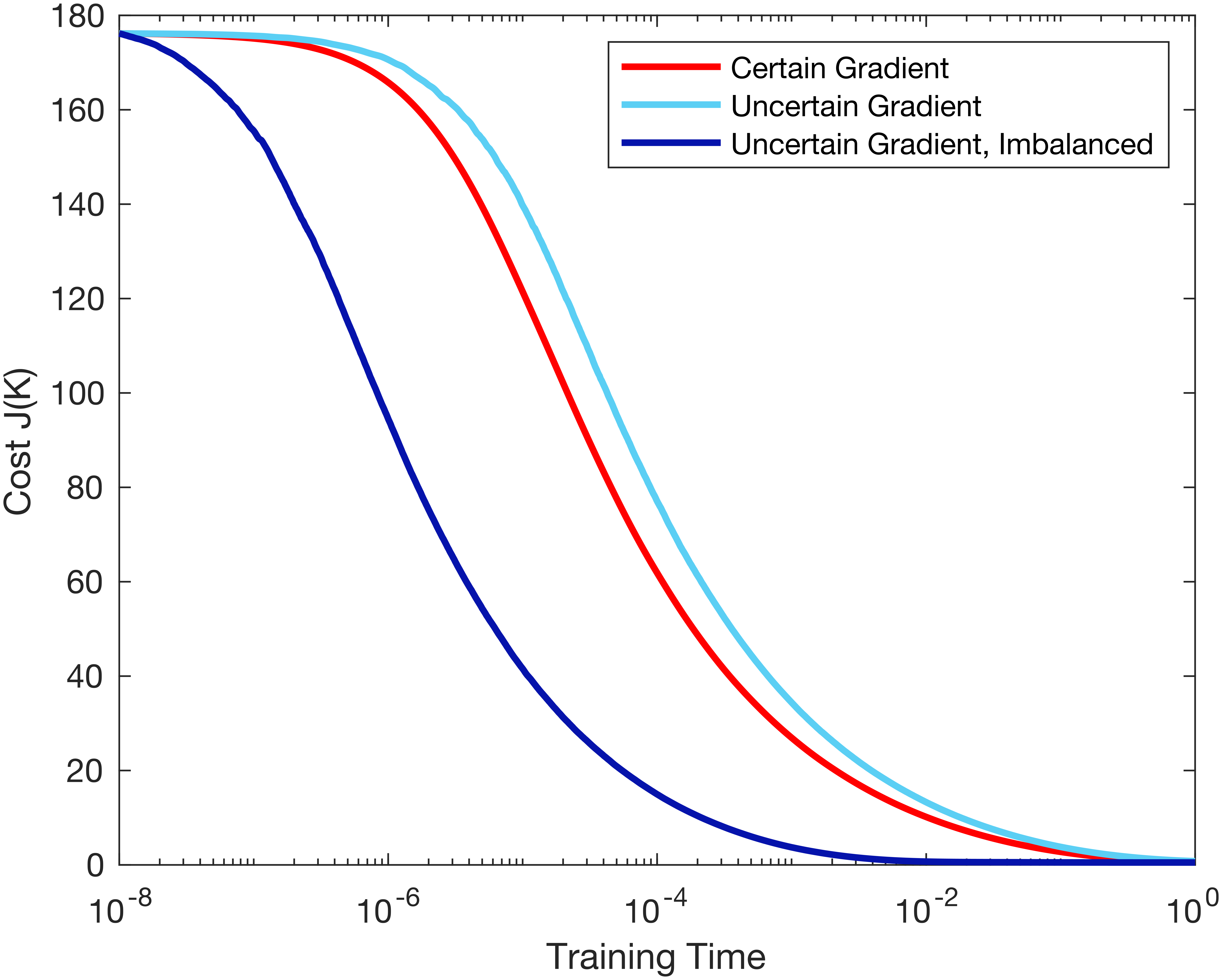}
    \caption{\small Simulation results for uncertain oracle. For some randomly picked initialization, the red curve shows the time evolution of the cost function for the vector field generated when the gradient if perfectly computed through its closed-form expression. In light and dark blue, the gradient is estimated numerically through evaluations of the cost function, with the light blue trajectory being the one initialized at the same point as the red trajectory, and the dark blue being the one initialized at the same point as the other two, except for an imbalance factor of 10, as described in Section \ref{sec:simul}.
    }
      \label{fig:SimulUncOrc}
     
\end{figure}

We now look at the case where the exact value of the gradient is unknown, and the algorithm samples the value of the cost function at different directions around the current point to estimate it numerically. The code for this set of simulations is also available at \cite{paperRepo2025}.

The gradient is estimated by disturbing the cost at the current value of $(K_1,K_2)$ in the direction of $20$ different elementary matrices, \ie in the direction of $20$ different entries of $(K_1,K_2)$. The resulting estimated gradient can be viewed as the true gradient plus a noise term. All simulations are done for the same initialization, picked randomly in a distribution around zero -- this works for our example because $A$ was specifically selected to be stable.

The resulting solutions are displayed in Fig. \ref{fig:SimulUncOrc}. Notice that the solution computed with perfect knowledge of the gradient and no enforced imbalance (in red) converges faster than the balanced initialization with the estimated gradient (light blue). However, once we increase the imbalance of the initialization by a factor of $10$, the resulting solution (dark blue) converges much quicker than even the solution without uncertainty. This indicates that the disturbance caused by the uncertainty in the dynamics can be overcome by the acceleration brought by imbalanced initializations.

\section{Conclusions}
\label{sec:conclusions}

This paper investigated the use of linear feedforward neural networks (LFFNNs) for computing the optimal solution of the LQR problem. The theoretical exploration conducted yielded several important results, as summarized below.

In Section \ref{sec:ThrBknd} we revised key literature results on both gradient methods for the LQR problem and for overparameterized linear regressions, both areas that compose the main contributions of this paper. Then, in Section \ref{sec:NNLQR} we introduced the overparameterized policy-optimization LQR problem (poLQR) and proved the main theoretical results of the paper regarding convergence of the solutions in Theorem \ref{thm:existence}. Also in this section, we deepened our analysis of the case with a single hidden layer, proving almost everywhere convergence to the optimal feedback matrix in Theorem \ref{thm:Saddles} and Corollary \ref{cor:AEConv2Opt}, and characterizing all saddles in Lemma \ref{lem:critcharact}. We believe these results serve as a strong basis from which to derive an intuitive understanding of the behavior of the solutions of overparameterized formulations. To better develop such intuition we proceeded in Section \ref{sec:scalarcase} with analyzing the vector case, whose simpler setup allows for explicit computation of convergence conditions to the different critical points of the problem. Then, in Section \ref{sec:simul}, we performed a comprehensive numerical analysis of the problem, showing how different initializations affect the convergence when compared to a non-overparameterized poLQR formulation. The simulations illustrate the distinct behavior the solution can present depending on its initialization and show how the overparameterized formulation can accelerate or deaccelerate the convergence of the solution to the optimal solution of the poLQR. The simulations indicate that a solution can be arbitrarily accelerated by increasing levels of imbalance for the initialization, however, the stiffness of the resulting ODE provides a practical trade-off to the acceleration.

Many open problems related to the work in this paper remain. A natural follow-up question is how general an optimization problem can be for an overparameterized formulation to hold the properties characterized in this paper. Alternatively, one might be interested in possible practical applications of properties observed in this work, in which case sample and computational complexity analysis are essential to rigorously establishing the trade-offs of adopting such approach in practice. A more specific open problem lies in the characterization of the center-stable manifold of the saddles of the overparameterized gradient flow. In this paper we indicated what we believe are the main obstacles to doing so; however, if that were to be done in a future work, it could be leveraged to state formal robustness results for the general case and improve the general understanding of the behavior of the solution.

\bibliographystyle{plain}        
\bibliography{NeuralNetLQR}           

\begin{thebibliography}{10}

\bibitem{agarwal2021theory}
Alekh Agarwal, Sham~M Kakade, Jason~D Lee, and Gaurav Mahajan.
\newblock On the theory of policy gradient methods: {Optimality}, approximation, and distribution shift.
\newblock {\em Journal of Machine Learning Research}, 22(98):1--76, 2021.

\bibitem{alali2023reinforcement}
Mohammad Alali and Mahdi Imani.
\newblock Reinforcement learning data-acquiring for causal inference of regulatory networks.
\newblock In {\em American Control Conference (ACC), IEEE}, 2023.

\bibitem{arora2019fine}
Sanjeev Arora, Simon Du, Wei Hu, Zhiyuan Li, and Ruosong Wang.
\newblock Fine-grained analysis of optimization and generalization for overparameterized two-layer neural networks.
\newblock In {\em International Conference on Machine Learning}, pages 322--332. PMLR, 2019.

\bibitem{bah_learning_2022}
Bubacarr Bah, Holger Rauhut, Ulrich Terstiege, and Michael Westdickenberg.
\newblock Learning deep linear neural networks: {Riemannian} gradient flows and convergence to global minimizers.
\newblock {\em Information and Inference: A Journal of the IMA}, 11(1):307--353, March 2022.

\bibitem{baldi_neural_1989}
Pierre Baldi and Kurt Hornik.
\newblock Neural networks and principal component analysis: {Learning} from examples without local minima.
\newblock {\em Neural Networks}, 2(1):53--58, January 1989.

\bibitem{chitour_geometric_2023}
Yacine Chitour, Zhenyu Liao, and Romain Couillet.
\newblock A geometric approach of gradient descent algorithms in linear neural networks.
\newblock {\em Mathematical Control and Related Fields}, 13(3):918--945, 2023.

\bibitem{colding2014lojasiewicz}
Tobias~Holck Colding and William~P Minicozzi~II.
\newblock Lojasiewicz inequalities and applications.
\newblock In Shing-Tung~Yau Huai-Dong~Cao, Richard~Schoen, editor, {\em XIX of Surveys in Differential Geometry}. International Press of Boston, Boston, 2014.

\bibitem{cui2024small}
Leilei Cui, Zhong-Ping Jiang, and Eduardo~D Sontag.
\newblock Small-disturbance input-to-state stability of perturbed gradient flows: {Applications} to {LQR} problem.
\newblock {\em Systems \& Control Letters}, 188:105804, 2024.

\bibitem{de2023dynamics}
Arthur Castello~B de~Oliveira, Milad Siami, and Eduardo~D Sontag.
\newblock Dynamics and {Perturbations} of {Overparameterized} {Linear} {Neural} {Networks}.
\newblock In {\em 2023 62nd IEEE Conference on Decision and Control (CDC)}, pages 7356--7361. IEEE, 2023.

\bibitem{de2024remarks}
Arthur Castello~B de~Oliveira, Milad Siami, and Eduardo~D Sontag.
\newblock Remarks on the {Gradient} {Training} of {Linear} {Neural} {Network} {Based} {Feedback} for the {LQR} {Problem}.
\newblock In {\em 2024 63rd IEEE Conference on Decision and Control (CDC)}, pages 7846--7852. IEEE, 2024.

\bibitem{paperRepo2025}
Arthur Castello~Branco de~Oliveira.
\newblock {OVP}\_{M}odel{F}ree\_{LQR}\_{A}utomatica2025.
\newblock \url{https://github.com/ArthurCBO/OVP_ModelFree_LQR_Automatica2025.git}, 2025.

\bibitem{eftekhari_training_2020}
Armin Eftekhari.
\newblock Training {Linear} {Neural} {Networks}: {Non}-{Local} {Convergence} and {Complexity} {Results}.
\newblock In {\em Proceedings of the 37th {International} {Conference} on {Machine} {Learning}}, pages 2836--2847. PMLR, November 2020.
\newblock ISSN: 2640-3498.

\bibitem{fazel2018global}
Maryam Fazel, Rong Ge, Sham Kakade, and Mehran Mesbahi.
\newblock Global convergence of policy gradient methods for the linear quadratic regulator.
\newblock In {\em International conference on machine learning}, pages 1467--1476. PMLR, 2018.

\bibitem{gravell2020learning}
Benjamin Gravell, Peyman~Mohajerin Esfahani, and Tyler Summers.
\newblock Learning optimal controllers for linear systems with multiplicative noise via policy gradient.
\newblock {\em IEEE Transactions on Automatic Control}, 66(11):5283--5298, 2020.

\bibitem{hu2023toward}
Bin Hu, Kaiqing Zhang, Na~Li, Mehran Mesbahi, Maryam Fazel, and Tamer Ba{\c{s}}ar.
\newblock Toward a theoretical foundation of policy optimization for learning control policies.
\newblock {\em Annual Review of Control, Robotics, and Autonomous Systems}, 6(1):123--158, 2023.

\bibitem{jin2017escape}
Chi Jin, Rong Ge, Praneeth Netrapalli, Sham~M Kakade, and Michael~I Jordan.
\newblock How to escape saddle points efficiently.
\newblock In {\em International conference on machine learning}, pages 1724--1732. PMLR, 2017.

\bibitem{kawaguchi_deep_2016}
Kenji Kawaguchi.
\newblock Deep {Learning} without {Poor} {Local} {Minima}.
\newblock In {\em Advances in {Neural} {Information} {Processing} {Systems}}, volume~29. Curran Associates, Inc., 2016.

\bibitem{kelley1975general}
J.L. Kelley.
\newblock {\em General {0}}.
\newblock Graduate Texts in Mathematics. Springer New York, 1975.

\bibitem{2016_lee_et_al_recht}
Jason~D. Lee, Max Simchowitz, Michael~I. Jordan, and Benjamin Recht.
\newblock Gradient {Descent} {Only} {Converges} to {Minimizers}.
\newblock In Vitaly Feldman, Alexander Rakhlin, and Ohad Shamir, editors, {\em 29th Annual Conference on Learning Theory}, volume~49 of {\em Proceedings of Machine Learning Research}, pages 1246--1257, Columbia University, New York, New York, USA, 23--26 Jun 2016. PMLR.

\bibitem{levine_determination_1970}
W.~Levine and M.~Athans.
\newblock On the determination of the optimal constant output feedback gains for linear multivariable systems.
\newblock {\em IEEE Transactions on Automatic Control}, 15(1):44--48, February 1970.

\bibitem{lojasiewicz1984gradients}
Stanis{\l}aw {\L}ojasiewicz.
\newblock Sur les trajectoires du gradient d'une fonction analytique. ({Trajectories} of the gradient of an analytic function).
\newblock {\em Semin. Geom., Univ. Studi Bologna}, 1982/1983:115--117, 1984.

\bibitem{min_explicit_2021}
Hancheng Min, Salma Tarmoun, Rene Vidal, and Enrique Mallada.
\newblock On the {Explicit} {Role} of {Initialization} on the {Convergence} and {Implicit} {Bias} of {Overparametrized} {Linear} {Networks}.
\newblock In {\em Proceedings of the 38th {International} {Conference} on {Machine} {Learning}}, pages 7760--7768. PMLR, July 2021.
\newblock ISSN: 2640-3498.

\bibitem{min2023convergence}
Hancheng Min, Ren{\'e} Vidal, and Enrique Mallada.
\newblock On the convergence of gradient flow on multi-layer linear models.
\newblock In {\em International Conference on Machine Learning}, pages 24850--24887. PMLR, 2023.

\bibitem{mohammadi2021lack}
Hesameddin Mohammadi, Mahdi Soltanolkotabi, and Mihailo~R Jovanovi{\'c}.
\newblock On the lack of gradient domination for linear quadratic {G}aussian problems with incomplete state information.
\newblock In {\em 2021 60th IEEE Conference on Decision and Control (CDC)}, pages 1120--1124. IEEE, 2021.

\bibitem{mohammadi_convergence_2022}
Hesameddin Mohammadi, Armin Zare, Mahdi Soltanolkotabi, and Mihailo~R. Jovanovic.
\newblock Convergence and {Sample} {Complexity} of {Gradient} {Methods} for the {Model}-{Free} {Linear}–{Quadratic} {Regulator} {Problem}.
\newblock {\em IEEE Transactions on Automatic Control}, 67(5):2435--2450, May 2022.

\bibitem{motamedi2024robustness}
Elaheh Motamedi, Kian Behzad, Rojin Zandi, Hojjat Salehinejad, and Milad Siami.
\newblock Robustness evaluation of machine learning models for robot arm action recognition in noisy environments.
\newblock In {\em ICASSP 2024-2024 IEEE International Conference on Acoustics, Speech and Signal Processing (ICASSP)}, pages 6215--6219. IEEE, 2024.

\bibitem{nesterov2006cubic}
Yurii Nesterov and Boris~T Polyak.
\newblock Cubic regularization of {N}ewton method and its global performance.
\newblock {\em Mathematical programming}, 108(1):177--205, 2006.

\bibitem{2017_panageas_et_al_gradient_descent}
Ioannis Panageas and Georgios Piliouras.
\newblock {Gradient {Descent} {Only} {Converges} to {Minimizers}: {Non}-{Isolated} {Critical} {Points} and {Invariant} {Regions}}.
\newblock In Christos~H. Papadimitriou, editor, {\em 8th Innovations in Theoretical Computer Science Conference (ITCS 2017)}, volume~67 of {\em Leibniz International Proceedings in Informatics (LIPIcs)}, pages 2:1--2:12, Dagstuhl, Germany, 2017. Schloss Dagstuhl -- Leibniz-Zentrum f{\"u}r Informatik.

\bibitem{polyak1963gradient}
Boris~T Polyak.
\newblock Gradient methods for the minimisation of functionals.
\newblock {\em USSR Computational Mathematics and Mathematical Physics}, 3(4):864--878, 1963.

\bibitem{rautert_computational_1997}
T.~Rautert and E.~W. Sachs.
\newblock Computational {Design} of {Optimal} {Output} {Feedback} {Controllers}.
\newblock {\em SIAM Journal on Optimization}, 7(3):837--852, August 1997.

\bibitem{ravari2022optimal}
Amirhossein Ravari, Seyede~Fatemeh Ghoreishi, and Mahdi Imani.
\newblock Optimal {Recursive} {Expert}-{Enabled} {Inference} in {Regulatory} {Networks}.
\newblock {\em IEEE Control Systems Letters}, 7:1027--1032, 2022.

\bibitem{ravari2024optimal}
Amirhossein Ravari, Seyede~Fatemeh Ghoreishi, and Mahdi Imani.
\newblock Optimal {Inference} of {Hidden} {M}arkov {Models} {Through} {Expert}-{Acquired} {Data}.
\newblock {\em IEEE Transactions on Artificial Intelligence}, 2024.

\bibitem{shub2013}
M.~Shub.
\newblock {\em Global {Stability} of {Dynamical} {Systems}}.
\newblock Springer, 2013.

\bibitem{sontag2013mathematical}
Eduardo~D Sontag.
\newblock {\em Mathematical control theory: deterministic finite dimensional systems}, volume~6.
\newblock Springer Science \& Business Media, 2013.

\bibitem{sontag_remarks_2022}
Eduardo~D. Sontag.
\newblock Remarks on input to state stability of perturbed gradient flows, motivated by model-free feedback control learning.
\newblock {\em Systems \& Control Letters}, 161:105138, March 2022.

\bibitem{sznaier2022role}
Mario Sznaier, Alex Olshevsky, and Eduardo~D Sontag.
\newblock The {Role} of {Systems} {Theory} in {Control} {Oriented} {Learning}.
\newblock In {\em 25th International Symposium on Mathematical Theory of Networks and Systems}, 2022.

\bibitem{tarmoun_understanding_2021}
Salma Tarmoun, Guilherme Franca, Benjamin~D. Haeffele, and Rene Vidal.
\newblock Understanding the {Dynamics} of {Gradient} {Flow} in {Overparameterized} {Linear} models.
\newblock In {\em Proceedings of the 38th {International} {Conference} on {Machine} {Learning}}, pages 10153--10161. PMLR, July 2021.
\newblock ISSN: 2640-3498.

\bibitem{wafi2023comparative}
Moh~Kamalul Wafi and Milad Siami.
\newblock A {Comparative} {Analysis} of {Reinforcement} {Learning} and {Adaptive} {Control} {Techniques} for {Linear} {Uncertain} {Systems}.
\newblock In {\em 2023 Proceedings of the Conference on Control and its Applications (CT)}, pages 25--32. SIAM, 2023.

\bibitem{zandi2023robot}
Rojin Zandi, Hojjat Salehinejad, Kian Behzad, Elaheh Motamedi, and Milad Siami.
\newblock Robot motion prediction by channel state information.
\newblock In {\em 2023 IEEE 33rd International Workshop on Machine Learning for Signal Processing (MLSP)}, pages 1--6. IEEE, 2023.

\bibitem{zhang2023revisiting}
Xiangyuan Zhang and Tamer Ba{\c{s}}ar.
\newblock Revisiting {LQR} control from the perspective of receding-horizon policy gradient.
\newblock {\em IEEE Control Systems Letters}, 7:1664--1669, 2023.

\end{thebibliography}

\appendix

\section{Systems with Strict Saddles}
    \label{apx:StrictSaddles}
    We state and prove a few more general results about the convergence of nonlinear systems with multiple equilibria. 

    In this section, we consider a general differential equation
    \be{eq:generic_system}
    \dot x=f(x)
    \ee
    evolving on an open subset $\statespace\subseteq \R^n$.
    We assume that $f:\statespace\rightarrow \R^n$ is continuously differentiable.
    The solution $x(t)=\phi (t,\initialx)$ of~\eqref{eq:generic_system} with initial state
    $\initialx\in \statespace$ is defined (and in $\statespace$) on a maximal interval
    $t\in (\Tmin,\Tmax)$, where $-\infty \leq \Tmin<0<\Tmax\leq +\infty $.
    The $n\times n$ Jacobian matrix of $f$ evaluated at a point $x\in \statespace$ is denoted
    by $\jac{f}(x)$.
    
    For any subset $S\subseteq \statespace$ define the finite-time domain of attraction
    $\domain(S)$ of $S$ as the set of all $\initialx\in \statespace$ such that
    $\Tmax=+\infty $ and there is some $\Thit\geq 0$ such that $\phi (t,\initialx)\in S$ for
    all $t\geq \Thit$. 
    
    We say that $\equilx\in \statespace$ is a \textit{strict saddle equilibrium}
    of~\eqref{eq:generic_system} if
    \ben
      \item
        $f(\equilx)=0$ and
      \item
        $\jac{f}(\equilx)$ has at least one eigenvalue with
      positive real part and at least one eigenvalue with non-positive real part.
    \een
    
    The following theorem and corollary generalize results for discrete-time
    gradient iterations that were given in \cite{2017_panageas_et_al_gradient_descent},
    which in turn generalized a result from~\cite{2016_lee_et_al_recht} that
    restricted to discrete sets of strict saddles.
    
    \begin{theorem}\label{th:convergence}
    Suppose that $\equilx\in \statespace$ is a strict saddle equilibrium
    of~\eqref{eq:generic_system}.
    Then there exists an open neighborhood $B\subseteq \statespace$ of $\equilx$ such
    that $\domain(B)$ has Lebesgue measure zero.
    \end{theorem}
    
    \noindent{\textbf{Proof.}}
    Pick any equilibrium point $\equilx\in \statespace$.
    Next modify the vector field $f$ to a vector field $g$ so that $g$ coincides with
    $f$ on an open neighborhood of $U$ of $\equilx$ and $g$ vanishes outside a
    compact set $K\subseteq \statespace$.
    Since $g$ has compact support, solutions are defined for all $t\in \R$, and the
    map $G: x\mapsto \gamma (1,x)$ (time-1 map for $g$, where $\gamma$ is the flow of $g$) is a $C^1$ diffeomorphism.
    Since $\gamma (1,\equilx)=\equilx$, it follows that $G(\equilx)=\equilx$, and since
    $G$ is a diffeomorphism, there is some neighborhood $V$ of $\equilx$ in which $G=F$,
    where $F$ is the time-1 map for $f$.
    The Center-Stable Manifold Theorem as, for example, stated in~\cite{shub2013},
    Theorem III.7, applied $G$ restricted to $V$, gives the existence of an open
    subset $B$ of $V$ and a local center stable manifold $W$ of dimension equal to
    the number of eigenvalues with nonpositive real part, with the property that
    for any $x\in B$ such that $G^\ell(x)\in V$ for all $\ell\in \Z_+$ necessarily $x\in W$. 
    Since $F=G$ on $V$, the same property is true for $F$.
    
    Pick any point $\initialx\in \domain(B)$ and pick $k=\Thit\geq 0$, without loss of
    generality a positive integer, such that $\phi (t,\initialx)\in B$ for all
    $t\geq k$. Let $x=\phi (k,\initialx)$.
    Then $F^\ell(x) = \phi (k+\ell,\initialx)\in B$ for all $\ell\in \Z_+$, and
    therefore necessarily $x\in W$.
    We have established that for each $\initialx\in \domain(B)$ there is some $k$ such
    that $F^k$, the time-$k$ map of the flow $f$, is defined at $\initialx$ and
    satisfies $F^k(\initialx)\in W$.
    It follows that $\domain(B)$ is the union of the (countably many) sets $S_k$
    consisting of those points $x\in \statespace$ such that $F^k(x)\in W$.
    Thus it will suffice to show that each set $S_k$ has Lebesgue measure zero.
    Note that $F^k$ is a local diffeomorphism, it being a time-$k$ map for a
    differentiable vector field. (It is not necessarily a global diffeomorphism, so
    we cannot argue that $(F^k)^{-1}(W)$ is diffeomorphic to $W$. In fact,
    preimages may not even belong to $\statespace$.)
    Thus, there is an open neighborhood $N_{\initialx}$ of $\initialx$ in $\statespace$
    that maps diffeomorphically by $F^k$ into an open neighborhood $M_{\initialx}$
    of $F^k(\initialx)$.
    By uniqueness of solutions in time $-k$, the preimage of $M_{\initialx}$ is
    exactly $N_{\initialx}$.
    Note that $S_k$ is included in the union $N_k$ over $\initialx\in\statespace$ of the sets $N_{\initialx}$.
    Also, for each $\initialx$, $N_{\initialx}\cap S_k$ maps diffeomorphically onto
    $M_{\initialx}\cap W$, and therefore $N_{\initialx}\cap S_k$ has Lebesgue measure zero (because $W$
    has measure zero and diffeomorphisms transform null sets into null sets). 
    Recall that Lindel\"of's Lemma (see e.g~\cite{kelley1975general})
    insures that every open cover of any subset $S$ of $\R^n$ (or more generally, of any
    second-countable space) admits a countable subcover.
    Applied to $N_k$, we have a countable subcover by sets $N_{\initialx_k}$, and for
    each of these $N_{\initialx_k}\cap S_k$ has measure zero, so $N_k\cap S_k = S_k$ has measure zero as well.
    \qed
    
    \begin{corollary}
    \label{cor:saddles}
    Suppose that $\equilset\subseteq \statespace$ is a set consisting of strict saddle
    equilibria of~\eqref{eq:generic_system}.
    Then the set $\convergingset$ of points $\initialx\in \statespace$ whose trajectories
    converge to points in $\equilset$ has measure zero.
    \end{corollary}  
    
    \noindent{\textbf{Proof.}}
    For each $\equilx\in \equilset$, we may pick by Theorem~\ref{th:convergence} an
    open neighborhood $B_{\equilx}\subseteq \statespace$ of $\equilx$ such that
    $\domain(B_{\equilx})$ has measure zero.
    The union of the sets $B_{\equilx}$ covers $\equilset$.
    By Lindel\"of's Lemma applied to $S = \equilset$, we conclude that there is a
    countable subset of balls $\{B_{\equilx_k},k\in \Z_{+}\}$ which covers $\equilset$.
    We claim that $\convergingset\subseteq \bigcup _k\domain(B_{\equilx_k})$.
    Since a union of measure zero sets has measure zero, this will establish the claim.
    So pick any $\initialx\in \convergingset$.
    Thus, $\phi (t,\initialx)\rightarrow \equilx$ for some $\equilx\in \equilset$.
    Since $\equilset\subseteq \bigcup _k B_{\equilx_k}$, it follows that
    $\equilx\in B_{\equilx_k}$ for some $k$.
    Since $B_{\equilx_k}$ is a neighborhood of $\equilx$, this means that
    there is some $\Thit\geq 0$ such that $\phi (t,\initialx)\in B_{\bar x_k}$ for all $t\geq \Thit$.
    Therefore $\initialx\in \domain(B_{\equilx_k})$.
    This completes the proof.
    \qed
    
%
\begin{corollary}
    \label{cor:AEConvf}
    Let $\loss$ be a real-analytic (loss) function from $\statespace$ into $\mathbb{R}_+$. Let $f$ be the gradient of $\loss$ and let the set $Z$ of points where $f(x)=0$ be the union of two sets $Z=M\cup S$, where $M$ is the set of points at which $\loss$ is minimized, and $S$ consists of strict saddles for the gradient flow dynamics:
    \begin{equation*}
        \dot x = -f(x).
    \end{equation*}
    Assume in addition that every trajectory of the gradient flow dynamics is pre-compact. Then, except for a set of measure zero, all trajectories converge to $M$. 
\end{corollary}

    \textbf{Proof:} Lojasiewcz's Theorem states that every pre-compact trajectory of a real analytic gradient system converges to a unique equilibrium. This theorem is given \cite{lojasiewicz1984gradients} and an excellent exposition is given in \cite{colding2014lojasiewicz}. From this, one can apply Corollary \ref{cor:saddles} and conclude that the set of initializations for which the trajectories of the system converge to $S$ must have measure zero. \qed
    
    %

\section{Proofs of Main Results}    
\label{apx:mainthms}

        \textbf{Proof of Theorem \ref{thm:jacCost}:} For any $i$ between $1$ and $N$, notice that under closed loop with $u = K_N\dots K_i\dots K_1 Cx$ the dynamics of the system become
        \begin{align*}
            \dot x &= (A+BK_n\dots K_i\dots K_1 C)x \\&= (A+B_iK_iC_i)x,
        \end{align*}
        which are equivalent to a system under simple output feedback $K_i$, input matrix $B_i$ and output matrix $C_i$. Now, since when computing the partial derivatives in \eqref{eq:GradFlowKi}, one fixes the value of all $K_j$ for $j\neq i$, computing $\partial J/\partial K_i$ is equivalent to computing the partial derivative for the linear system with parameter matrices $A$, $B_i$, $C_i$ and single feedback matrix $K_i$, as done in \eqref{eq:gradJLQR}.

        \textcolor{red!0!black}{From here, the remainder of the proof is obtained by applying Theorem 3.2 of \cite{rautert_computational_1997} to the system $(A,B_i, C_i, K_i)$, however we include it here for completeness.}

        Notice that for the system $(A,B_i,C_i, K_i)$, the LQ cost can be written as $J_{K_i}:=J(K_i) = \trace{P_{K_i}\Sigma_0}$. Furthermore, by definition $\nabla_{K_i}P_{K_i}$ is such that  one can write
        \begin{equation*}
            P_{K_i+\dd K_i} = P_{K_i}+\left\langle\nabla_{K_i}P_{K_i} \dd K_i\right\rangle + \text{h.o.t.},
        \end{equation*}
        where $\langle\cdot,\cdot\rangle$ is the appropriate inner product, which for a matrix space is $\langle A,B\rangle = A^\top B$. From here write
        \begin{align*}
            J_{K_i+\dd K_i} &= \trace{P_{K_i+\dd K_i}\Sigma_0} \\ &= \trace{(P_{K_i}+\nabla_{K_i}P_{K_i}^\top \dd K_i + \text{h.o.t})\Sigma_0} \\ &= J_{K_i} + \trace{\nabla_{K_i}P_{K_i}\dd K_i\Sigma_0} + \text{h.o.t}
        \end{align*}
        which implies that 
        $$\langle \nabla_{K_i}J_{K_i},\dd K_i\rangle = \trace{\nabla_{K_i}P_{K_i}\dd K_i\Sigma_0},$$
        by collecting the linear terms in $\dd K_i$. To compute $\nabla_{K_i}P_{K_i}\dd K_i$, consider \eqref{eq:LyapPtrad} applied to $P_{K_i+\dd K_i}$
        \begin{align*}
            &(P_{K_i}+\nabla_{K_i}P_{K_i}^\top \dd K_i + \text{h.o.t.})\\&\times (A+B_i(K_i+\dd K_i)C_i) \\&+ (A+B_i(K_i+\dd K_i)C_i)^\top \\&\times (P_{K_i}+\nabla_{K_i}P_{K_i}^\top \dd K_i + \text{h.o.t.}) &\nonumber \\&+ C_i^\top (K_i+\dd K_i)^\top R_i(K_i+\dd K_i)C_i + Q = 0.
        \end{align*}
        By expanding the equation above and collecting the linear terms one obtains the following matrix equality for $\nabla_{K_i}P_{K_i}^\top \dd K_i$
        \begin{align}
            &\nabla_{K_i}P_{K_i}^\top \dd K_i(A+B_iK_iC_i) \nonumber\\&+ (A+B_iK_iC_i)^\top \nabla_{K_i}P_{K_i}^\top \nonumber\\
            =&-C_i^\top \dd K_i^\top(B_i^\top P_{K_i}+R_iK_iC_i) \nonumber\\&- (B_i^\top P_{K_i}+R_iK_iC_i)^\top \dd K_i C_i \label{eq:LyapGradPk}
        \end{align}
        Then, multiply \eqref{eq:LyapGradPk} by $L_{K_i}$ and \eqref{eq:LyapLtrad} by $\nabla_{K_i}P_{K_i}$, take the trace of both equations and combine them to obtain
        \begin{align*}
            &\trace{\nabla_{K_i}P_{K_i}\dd K_i\Sigma_0} = \langle\nabla_{K_i}J_{K_i},\dd K_i\rangle \\&= \trace{2\left((B_i^\top P_{K_i}+R_iK_iC_i)L_{K_i}C_i^\top\right)^\top \dd K_i},
        \end{align*}
        completing the proof. \qed
        %
%

        \textbf{Proof of Theorem \ref{thm:invariance}:} First, notice from the definitions in Theorem \ref{thm:jacCost} that for all $i$ between $1$ and $N-1$
        \begin{align}
            B_i &= BK_N\dots K_{i+1} = B_{i+1}K_{i+1} \\
            C_{i+1}&= K_i\dots K_1 = K_iC_i \\
            R_i &= K_{i+1}^\top \dots K_N^\top R K_N \dots K_{i+1} \nonumber \\&= K_{i+1}^\top R_{i+1}K_{i+1}.
        \end{align}
        Then, from
        \begin{equation}
            \frac{d}{dt}(K_iK_i^\top) = \dot K_i K_i^\top + K_i\dot K_i^\top,
        \end{equation}
        notice that
        {\small \begin{align}
            &\dot K_iK_i^\top = -2(B_i^\top P_\K + R_iK_iC_i)L_\K C_i^\top K_i^\top\nonumber  \\ &= -2(K_{i+1}^\top B_{i+1}^\top P_\K+K_{i+1}^\top R_{i+1}K_{i+1}C_{i+1})L_\K C_{i+1}^\top \nonumber \\ &=K_{i+1}^\top 2(B_{i+1}^\top P_\K+R_{i+1}K_{i+1}C_{i+1})L_\K C_{i+1}^\top \nonumber \\ &= K_{i+1}^\top \dot K_{i+1},
        \end{align}}
        and similarly, $K_i\dot K_i^\top = \dot K_{i+1}^\top K_{i+1}$. With this, we write
        \begin{align*}
            \frac{d}{dt}(K_iK_i^\top - K_{i+1}^\top K_{i+1}) &= \dot K_iK_i^\top + K_i\dot K_i^\top \\&~~~~- \dot K_{i+1}^\top K_{i+1}-K_{i+1}^\top \dot K_{i+1} \\ &=0
        \end{align*}
        which proves the theorem. \qed

        \textbf{Proof of Theorem \ref{thm:existence}:} Begin the proof by establishing that under gradient flow, the cost function is non-increasing. This can be easily verified by computing
        \begin{equation}
            \dot{J} = \sum_{k=1}^N\left\langle \nabla_{K_k}J, \dot K_k\right\rangle = -\sum_{k=1}^N\|\nabla_{K_k}J\|_F^2\leq 0,
        \end{equation}
        where $\|\cdot \|_F$ is the Frobenius norm of a matrix. Notice that this proves that $P_\K$ is also non-increasing since $J = \trace{P_\K}$ is non-increasing (remember, it is assumed that $\Sigma_0=I$ although the proof would follow as is for any $\Sigma_0$ \sout{full rank}\highlight{positive definite}). The next step is to prove that for any trajectory, the parameter matrices $K_i$s are contained in some compact set. To do that, notice that since $J(\mathbf{K})\leq J(\mathbf{K})|_{t=0}$, $\tr(P_\mathbf{K})\leq \tr(P_\mathbf{K})|_{t=0}$. With this, and using the fact that for any $\mathbf{K}\in\mathcal{K}$, $P_\mathbf{K}$ is the solution of \eqref{eq:lyapP} which, defining $\overline K := K_NK_{N-1}\dots K_1$, is equivalent to saying that for any $x$ in the state space
        \begin{equation}
        \label{eqn:tocontradict}
        2x^\top(A+B\overline K)^\top P_\K x + x^\top (\overline K^\top R\overline K + Q)x = 0.
        \end{equation}
        This equality is then used to show that $\overline K$ lies in a compact by contradiction. To do that assume $\overline K$ is unbounded. This means that there exist a sequence of times $t_i$, $i=1,2,\dots$ such that $t_i>t_j$ for any $i>j$, and that for any $i$, there are two unitary vectors $x_i$ and $u_i$ of appropriate dimensions such that $\overline K(t_i) x_i = \lambda_i u_i$, with $\lambda_i>0$ and so that the sequence $\lambda_i\rightarrow\infty$. Substitute this into (\ref{eqn:tocontradict}) (omitting time dependencies for clarity), which results in:
        \begin{align*}
        &0\;=\;2x_i^\top(A+B\overline K)^\top P_\K x_i + x_i^\top (\overline K^\top R\overline K + Q)x_i \\
        &=\;2x_i^\top A^\top P_\K x_i + 2\lambda_i u_i^\top B^\top P_\K x_i \\&+ \lambda_i^2u_i^\top R u_i + x_i^\top Q x_i.
        \end{align*}
        This last expression is a quadratic function of $\lambda_i$ with leading coefficient $u_i^\top R u_i$ bounded below by $\sigma$, where $\sigma>0$ is the smallest eigenvalue of the positive definite matrix $R$. Thus, the leading term is bounded below by $\sigma\lambda_i^2$. The remaining terms are bounded by an expression $c\lambda_i$, where $c$ is an upper bound on $\|B^\top P_\K x_t\|$, which is finite because $P_\K$ is non-increasing, $B$ is assumed to be a finite parameter matrix of the system, and $x_i$ is unitary by definition.
        
        Define 
        \begin{align*}
            F(\lambda_i) &= 2x_i^\top A^\top P_\K x_i + 2\lambda_i u_i^\top B^\top P_\K x_i \\&~~~+ \lambda_i^2u_i^\top R u_i + x_i^\top Q x_i.
        \end{align*}
        It was remarked above that $F(\lambda_i) = 0$ for all $i$ 
        So also $F(\lambda_i)/\lambda_i^2=0$, since $\lambda_i>0$. However, 
        \begin{equation*}
            \lim_{i\rightarrow\infty} \frac{F(\lambda_i)}{\lambda_i^2} = \lim_{i\rightarrow\infty}u_i^\top R u_i \geq \sigma >0,
        \end{equation*}
        from which a contradiction is reached.
        
        Therefore, $\overline K$ is bounded within any trajectory. The next step is to show that $\overline K$ being bounded implies that all $K_i$s are also bounded. This part of the argument follows closely the proof of Proposition 1 in \cite{chitour_geometric_2023}, and of Theorem 3.2 of \cite{bah_learning_2022}, albeit done for the LQ cost rather than linear regression. 

        To show that $\overline K$ bounded implies $K_i$ bounded for all $i$, notice that a consequence of Theorem \ref{thm:invariance} is that for any $i,j$ between $1$ and $N$, there exist some constant $c_{ij}$ such that
        \begin{equation}
            \label{eq:normrelation}
            \|K_i\|_F^2 = \|K_j\|_F^2+\trace{W_{ij}}.
        \end{equation}
        where $W_{ij} = \sum_{k=i}^j{C_k}$ if $i<j$, $W_{ij} = -\sum_{k=j}^i{C_k}$ if $i>j$ and $W_{ij}=0$ if $i=j$. This is easily verified by taking the trace on both sides of \eqref{eq:invexp} and concatenating the resulting equations from $i$ to $j$. With this established, following relationship between $\overline K$ and $K_i$ can be used:
        \begin{equation}
            \label{eq:relKbarKi}
             \|K_i\|_F \leq \eta_i\|\overline K\|_F^{1/N}+\xi_i,
        \end{equation}
        where $\eta_i$ and $\xi_i$ depend only on the initialization of the parameter matrices. \textcolor{red!0!black}{The derivation of \eqref{eq:relKbarKi} is briefly given below for completeness, however, it can be obtained in the same way as equation (3.1) of \cite{bah_learning_2022} (from where this part of this proof is based on) once \eqref{eq:normrelation} is established, regardless of the different cost functions.} To derive \eqref{eq:relKbarKi}, consider \eqref{eq:invexp} iteratively to obtain
        \begin{align*}
            \overline K \overline K^\top &= K_N\dots K_1K_1^\top \dots K_N^\top \\
            &= K_N\dots K_2(\mathcal{C}_1+K_2^\top K_2)K_2^\top \dots K_N^\top \\ &= (K_NK_N^\top)^N+\mathbf{P}(K_2,\dots, K_N),
        \end{align*}
        where $\mathbf{P}(K_2,\dots,K_N)$ is a polynomial on the various $K_i$ and their transpose, whose degree is at most $2N-2$. Let $\sigma_N$ be the largest singular value of $K_N$, then
        \begin{align*}
            \sigma_N^{2N}&\leq \|(K_NK_N^\top)^N\|_F 
            \\ &~~~~~~~~~~\cdots \\ &\leq \|\overline{K}\;\overline{K}^\top\|^N_F+\|\mathbf{P}(K_2, K_N)\|_F.
        \end{align*}
        From \eqref{eq:normrelation}, we know that $\|K_i\|_F$ and $\|K_N\|_F$ differ only by a constant. Therefore there exist constants $a_i$ and $b_i$ such that $\|K_i\|_F\leq a_i\sigma_N+b_i$ for all $i$ between $1$ and $N$. From this, it follows that
        \begin{equation*}
            \|\mathbf{P}(K_2,\dots,K_N)\|_F\leq \mathbf{P}_N(\sigma_N),
        \end{equation*}
        where $\mathbf{P}_N$ is some monovariable polynomial of degree at most $2N-2$. One can always find some constant $\mathbf{C}$ such that $|\mathbf{P}_N(x)|\leq 0.5x^{2N}+\mathbf{C}$ for all $x>0$. Therefore, we can write
        \begin{align*}
            \sigma_N^{2N}&\leq \|(K_NK_N^\top)^N\|_F 
            \\ &\leq \|KK^\top\|_F+\|\mathbf{P}(K_2,\dot,K_N) \\ &\leq \|KK^\top\|_F+\mathbf{P}_N(\sigma_N) \\ &\leq \|KK^\top\|_F+ 0.5\sigma_N^{2N}+\mathbf{C}
        \end{align*}
        which implies that
        \begin{equation*}
            \sigma_N\leq \mathbf{B}_N\|K\|_F^{1/N}+\tilde {\mathbf{B}}_N,
        \end{equation*}
        for some constants $\mathbf{B}_N$ and $\tilde{\mathbf{B}}_N$. One can then finish the proof of \eqref{eq:normrelation} by using the relation that $\|K_i\|_F\leq a_i\sigma_N+b_i$.
         
        Therefore, once \eqref{eq:relKbarKi} is established, one can conclude that if $\overline K$ is bounded, then all $K_i$ also are. 

        Knowing that the $K_i$s are the solutions of a gradient system for a system with an analytic loss function (analyticity can be proved by an implicit function argument) and that these solutions remain in a compact for all time, the remainder of the proof is an application of Lojasiewicz's Theorem \cite{chitour_geometric_2023}. \qed
%
%

        \textbf{Proof of Theorem \ref{thm:Saddles}:} We break this proof into smaller steps. First we characterize the Hessian function of the cost function \eqref{eq:JOVPLQR} by collecting the second order terms of its Taylor expansion  as follows
        \begin{align*}
            &J((K_2+\dd K_2)(K_1+\dd K_1)) = \\&~~~~J(K_2K_1)+ \nabla_{K_1}J(K_2K_1)\dd K_1\\&~~~~+ \nabla_{K_2}J(K_2K_1)\dd K_2 \\&~~~~\frac{1}{2}\nabla^2_{K_1^2}J(K_2K_1)\dd K_1^2+\frac{1}{2}\nabla^2_{K_2^2}J(K_2K_1)\dd K_2^2 \\ &~~~~ \nabla^2_{K_1 K_2}J(K_2K_1)\dd K_1\dd K_2 +\text{h.o.t.}\\ &=J(K_2K_1)+J'_{\dd K_1}(K_2K_1)+J'_{\dd K_2}(K_2K_1)\\&~~~+\frac{1}{2}J''_{\dd K_1^2}(K_2K_1)+\frac{1}{2}J''_{\dd K_2^2}(K_2K_1)\\&~~~+J''_{\dd K_1 \dd K_2}(K_2K_1)+\text{h.o.t.}
        \end{align*}
        The expression for $J''_{dK_1^2}(K_2K_1)$ (and similarly for $J''_{dK_2^2}$) can be derived by looking at the Taylor expansion of $J'_{\dd K_1}(K_2(K_1+dK_1))$ as follows
        \begin{align*}
            &J'_{\dd K_1}(K_2(K_1+\dd K_1)) = \\
            & \tr(\dd K_1^\top K_2^\top 2[BP_\K+BP'_{\dd K_1}+RK_2K_1\\&~~~~~~+RK_2\dd K_1](L_\K+L'_{\dd K_1})) \\
            &= J'_{\dd K_1}(K_2K_1)\\&+\trace{\dd K_1^\top K_2^\top 2[BP_\K+RK_2K_1]L'_{\dd K_1}} \\
            &+\trace{\dd K_1^\top K_2^\top 2[BP'_{\dd K_1}+RK_2\dd K_1]L_\K}\\&+\text{h.o.t.},
        \end{align*}
        resulting in
        \begin{align*}
            &J''_{\dd K_1^2}(K_2K_1) \\&= \trace{\dd K_1^\top K_2^\top 2[BP_\K+RK_2K_1]L'_{\dd K_1}} \\
            &+\trace{\dd K_1^\top K_2^\top 2[BP'_{\dd K_1}+RK_2\dd K_1]L_\K},
        \end{align*}
        where $P'_{\dd K_1}$ and $L'_{\dd K_1}$ solve the following Lyapunov Equations respectively
        \begin{align}
            &P'_{\dd K_1}[A+BK_2K_1]+[A+BK_2K_1]^\top P'_{\dd K_1} = \nonumber\\&-\dd K_1^\top K_2^\top [B^\top P_\K+RK_2K_1]\label{eq:LyapGradPK1}\\&-[B^\top P_\K+RK_2K_1]^\top K_2\dd K_1\nonumber \\ 
            &L'_{\dd K_1}[A+BK_2K_1]^\top + [A+BK_2K_1]L'_{\dd K_1} = \nonumber\\&-BK_2\dd K_1L_\K-L_\K \dd K_1^\top K_2^\top B^\top \label{eq:LyapGradLK1}.
        \end{align}
        Multiplying \eqref{eq:LyapGradPK1} by $L'_{\dd K_1}$ and \eqref{eq:LyapGradLK1} by $P'_{\dd K_1}$ and taking the trace of both results in the following relation
        \begin{align*}
            &2\trace{dK_1^\top K_2^\top [B^\top P_\K+RK_2K_1]L'_{\dd K_1}} \\&= 2\trace{dK_1^\top K_2^\top [BP'_{\dd K_1}]L_\K},
        \end{align*}
        which allows us to simplify $J''_{\dd K_1^2}$ to
        \begin{align*}
            J''_{\dd K_1^2}(K_2K_1) &= 4\trace{\dd K_1^\top K_2^\top BP'_{\dd K_1}L_\K} \\
            &+2\trace{\dd K_1^\top K_2^\top RK_2\dd K_1L_\K}.
        \end{align*}
        Analogously, one can compute
        \begin{align*}
            J''_{\dd K_2^2}(K_2K_1) &= 4\trace{K_1^\top \dd K_2^\top BP'_{\dd K_2}L_\K} \\
            &+2\trace{K_1^\top dK_2^\top RK_2K_1L_\K}.
        \end{align*}
        For computing $J''_{\dd K_1\dd K_2}(K_2K_1)$ one can look at the Taylor expansion of either $J'_{\dd K_1}((K_d+\dd K_2)K_1)$ or of $J'_{\dd K_2}(K_2(K_1+\dd K_1))$. Expanding $J'_{\dd K_2}(K_2(K_1+\dd K_1))$ results in
        \begin{align*}
            &J'_{\dd K_2}((K_2(K_1+\dd K_1)) = \\ &= \trace{2[B^\top P_\K+RK_2K_1]L_\K K_1^\top \dd K_2^\top}\\&+\trace{2[B^\top P'_{\dd K_1}+RK_2\dd K_1]L_\K K_1^\top \dd K_2^\top}\\&+\trace{2[B^\top P_\K + RK_2K_1]L'_{\dd K_1}K_1^\top \dd K_2^\top}\\ &+ \trace{2[B^\top P_\K + RK_2K_1]L_\K \dd K_1^\top \dd K_2^\top} \\ &+\text{h.o.t.},
        \end{align*}
        which allows the conclusion that
        \begin{align*}
            &J''_{\dd K_1 \dd K_2}(K_2K_1) = \\&+\trace{2[B^\top P'_{\dd K_1}+RK_2\dd K_1]L_\K K_1^\top \dd K_2^\top}\\&+\trace{2[B^\top P_\K + RK_2K_1]L'_{\dd K_1}K_1^\top \dd K_2^\top}\\ &+ \trace{2[B^\top P_\K + RK_2K_1]L_\K \dd K_1^\top \dd K_2^\top}.
        \end{align*}
        Similarly, by expanding $J'_{\dd K_1}((K_2+\dd K_2)K_1)$ one can obtain that 
        \begin{align*}
            &J''_{\dd K_2 \dd K_1}(K_2K_1) = \\&+\trace{2[B^\top P'_{\dd K_2}+R\dd K_2K_1]L_\K \dd K_1^\top K_2^\top}\\&+\trace{2[B^\top P_\K + RK_2K_1]L'_{\dd K_2}\dd K_1^\top K_2^\top}\\ &+ \trace{2[B^\top P_\K + RK_2K_1]L_\K \dd K_1^\top \dd K_2^\top}.
        \end{align*}
        and notice that while $P'_{\dd K_1}$ and $L'_{\dd K_1}$ solve \eqref{eq:LyapGradPK1} and $\eqref{eq:LyapGradLK1}$, $P'_{\dd K_2}$ and $L'_{\dd K_2}$ respectivelly solve the following two Lyapunov Equations
        \begin{align}
            &P'_{\dd K_2}[A+BK_2K_1]+[A+BK_2K_1]^\top P'_{\dd K_2} = \nonumber\\&-K_1^\top \dd K_2^\top [B^\top P_\K+RK_2K_1]\label{eq:LyapGradPK2}\\&-[B^\top P_\K+RK_2K_1]^\top \dd K_2 K_1\nonumber \\ 
            &L'_{\dd K_2}[A+BK_2K_1]^\top + [A+BK_2K_1]L'_{\dd K_2} = \nonumber\\&-B\dd K_2K_1L_\K-L_\K K_1^\top \dd K_2^\top B^\top \label{eq:LyapGradLK2}.
        \end{align}
        Finally, by multiplying \eqref{eq:LyapGradPK1} and \eqref{eq:LyapGradLK2} by $L'_{\dd K_2}$ and by $P'_{\dd K_1}$ respectively and taking the trace, one reaches the following equation
        \begin{align*}
            &\trace{L'_{\dd K_2}dK_1^\top K_2^\top [B^\top P_\K+RK_2K_1]}\\&= \trace{P'_{\dd K_1}L_\K K_1^\top \dd K_2 B^\top},
        \end{align*}
        which allows the simplification of $J''_{\dd K_2\dd K_1}$ into
        \begin{align*}
            &J''_{\dd K_2 \dd K_1} =  \\&+\trace{2[B^\top P'_{\dd K_2}+R\dd K_2K_1]L_\K \dd K_1^\top K_2^\top}\\&+\trace{2P'_{\dd K_1}L_\K K_1^\top \dd K_2 B^\top}\\ &+ \trace{2[B^\top P_\K + RK_2K_1]L_\K \dd K_1^\top \dd K_2^\top}.
        \end{align*}
        With this, we have collected all second order terms of the Taylor expansion into the Hessian functon defined as
        \begin{align}
            &H(K_1, K_2, \dd K_1, \dd K_2) := \frac{1}{2}J''_{\dd K_1^2}(K_1, K_2)\nonumber\\&+\frac{1}{2}J''_{\dd K_2^2}(K_1, K_2)+J''_{\dd K_1 \dd K_2}(K_1, K_2)\label{eq:Hessian}.
        \end{align}
%
%

        We now move on with proving the theorem itself Let $(K_1,K_2)$ be a critical point of the gradient flow, that is, be such that
        \begin{align*}
            \dot K_1 &= \nabla_K J(K_2K_1)K_2^\top = 0 \\
            \dot K_2 &= K_1^\top \nabla_K J(K_2K_1) = 0,
        \end{align*}
        \highlight{where $\nabla_KJ(K_2K_1)$ is the expression in \eqref{eq:gradJLQR} computed for $K=K_2K_1$.} It is known \cite{rautert_computational_1997} that for the non-overparameterized gradient, $\nabla_KJ(K)=0$ and $K\in\mathcal{K}$ only if $K=K^*$, the optimal solution to the LQR problem. \highlight{When overparameterzing the problem,} however, the conditions above hold for any $K_1$, $K_2$ orthogonal to $\nabla_KJ(K_2K_1)$, even if $\nabla_KJ(K_2K_1)\neq0$.

        This orthogonality implies that there must exist two unitary vectors $\psi$ and $\phi$ such that $K_1^\top \phi = 0$, $\psi^\top K_2^\top = 0$, $\phi^\top \nabla_K J(K_2K_1) = \lambda \psi$ and $\nabla_KJ(K_2K_1)\psi = \lambda\phi^\top$ for some $\lambda<0$, for if no such vectors existed and $\nabla_KJ(K_2K_1)\neq 0$ then $\dot K_1=0$ and $\dot K_2=0$ could never hold. To prove this by contradiction, assume no such $\psi$ and $\phi$ exist, and since $\nabla_KJ(K_2,K_1)\neq0$ by assumption, let $\psi$ and $\phi$ be any unitary vectors such that $\nabla_KJ(K_2K_1)\psi = \lambda\phi^\top$ for some $\lambda$. Then notice that $\dot K_2\psi = K_1^\top \nabla_K J(K_2K_1)\psi = K_1^\top \phi\lambda \neq 0$ and $\phi^\top \dot K_1 = \phi^\top \nabla_KJ(K_2K_1)K_2^\top = \lambda\psi^\top K_2^\top \neq 0$, reaching contradiction.

        With this, let $\gamma_1$ and $\gamma_2$ be any two unitary vectors such that $\gamma_1^\top K_1^\top = 0$, $K_2^\top\gamma_2=0$\highlight{, and $\gamma_1^\top \gamma_2>0$ (if $\gamma_1^\top\gamma_2<0$, simply pick $-\gamma_1$ instead). D}efine $\dd K_1 = \psi\gamma_2^\top$ and $\dd K_2 = \phi \gamma_1^\top$. For this choice of $\dd K_1$ and $\dd K_2$ notice that $\dd K_2 K_1 = 0$ and $K_2 \dd K_1 = 0$, which implies that $J''_{dK_1^2} = 0$, $J''_{dK_2^2} = 0$, and
        \begin{align*}
            &J''_{dK_1dK_2}(K_2K_1) = \\& \trace{\dd K_2^\top \nabla_KJ(K_2K_1)\dd K_1^\top} \leq \lambda <0,
        \end{align*}
        proving that the Hessian has at least one negative eigenvalue, which implies that the spurious equilibria is a strict saddle of the gradient flow. \qed
        %
        %
        
        \highlight{
        Before proceeding to prove Lemma \ref{lem:critcharact}, following the order in the paper, we must first introduce and prove an auxiliary lemma that will help us prove Lemma \ref{lem:critcharact}.

        \begin{lemma}
            \label{lem:aux1}
            Given two matrices $A\in\mathbb{R}^{p\times o}$ and $B\in\mathbb{R}^{q\times o}$ for $p,q,o\in\mathbb{N}$ with $q\geq o$, the following two statements are equivalent
                
            \begin{enumerate}
                \item $AB^\top=0$;
                \item {There exist orthogonal matrices $\Psi_A$, $\Phi$, and $\Psi_B$, and rectangular diagonal matrices with non-negative diagonal elements
            $\Sigma_A$ and $\Sigma_B$, such that
            \begin{equation}
                \label{eq:auxlemA}
                \begin{split}
                    &A = \Psi_A\Sigma_A\Phi^\top
                \end{split}
            \end{equation}
            and
            \begin{equation}
                \label{eq:auxlemB}
                B = \Psi_B\Sigma_B\Phi^\top
            \end{equation}
            are SVDs of $A$ and $B$, and $\Sigma_A\Sigma_B^\top =0$.}
            \end{enumerate}
            
            Furthermore, in $2)$ we can write $\Sigma_A$ and $\Sigma_B$ as
    
            \begin{equation*}
                \Sigma_A = \begin{bmatrix}
                    \bar \Sigma_A & 0 & 0 \\ 0 & 0 & 0 \\ 0 & 0 & 0
                \end{bmatrix}
            \end{equation*}
            and
            \begin{equation*}
                \Sigma_B =\begin{bmatrix}
                    0 & 0 & 0 \\ 0 & \bar\Sigma_B & 0 \\ 0 & 0 & 0
                \end{bmatrix}
            \end{equation*}
            where $\bar\Sigma_A$ and $\bar\Sigma_B$ are diagonal matrices whose main diagonal elements are the nonzero singular values of $A$ and $B$ respectively.
        \end{lemma}

        \textbf{Proof of Lemma \ref{lem:aux1}:} The proof of $2)$$\Rightarrow$$1)$ follows immediately from writing
            
            $$AB^\top = \Psi_A\Sigma_A\Phi^\top\Phi\Sigma_B^\top \Psi_B^\top = \Psi_A\Sigma_A\Sigma_B^\top\Psi_B^\top = 0.$$
            
            To prove $1)$$\Rightarrow$$2)$, let $a = \rank{A}$ and $b=\rank{B}$, then write $A$ and $B$ as a sum of rank one matrices as follows
            \begin{equation}
                \label{eq:svdA}
                A = \sum_{i=1}^a \psi_{i,A}\phi_{i,A}^\top\sigma_{i,A}
            \end{equation}
            and
            \begin{equation}
                \label{eq:svdB}
                B = \sum_{i=1}^b\psi_{i,B}\phi_{i,B}^\top \sigma_{i,B},
            \end{equation}
            where $\psi_{i,A/B}$ and $\phi_{i,A/B}$ are left and right singular vectors of $A$ and $B$ associated with nonzero singular values, and $\sigma_{i,A/B}$ are the nonzero singular values of $A$ and $B$.

            Notice that $AB^\top=0$ if and only if any vector in the span of the columns of $B^\top$ belongs to the kernel of $A$. Therefore, since $\colspan{B^\top} = \spanv{\{\phi_{i,B}\}_{i=1}^b}$ and $\kernel{A}$ is the orthogonal complement of $\spanv{\{\phi_{i,A}\}_{i=1}^a}$, which is equivalent to $\phi_{i,A}^\top\phi_{j,B}=0$ for all $(i,j)\in \{1,\dots,a\}\times\{1,\dots,b\}$, \ie $$\spanv{\{\phi_{i,A}\}_{i=1}^a}\perp\spanv{\{\phi_{i,B}\}_{i=1}^b}$$. 
            
            This immediately implies that $a+b\leq o$. Let $\bar o = o-a-b\geq 0$, then consider any set of orthonormal vectors $\{\phi_{i,0}\}_{i=1}^{\bar o}$ that completes an orthonormal basis of $\mathbb{R}^o$ from $\{\phi_{i,B}\}_{i=1}^b\cup\{\phi_{i,A}\}_{i=1}^a$. 
            
            Define $\Phi_A$, $\Phi_B$ and $\Phi_0$ as the matrices whose columns are the vectors of $\{\phi_{i,A}\}_{i=1}^a$, $\{\phi_{i,B}\}_{i=1}^b$, and $\{\phi_{i,0}\}_{i=1}^{\bar o}$, respectively. With these definitions, we can express the matrix $\Phi$ as
            \begin{equation*}
                \Phi = \begin{bmatrix}
                    \Phi_A & \Phi_B & \Phi_0
                \end{bmatrix}.
            \end{equation*}
            Next, consider the sets $\{\psi_{i,A}\}_{i=1}^{p}$ and $\{\psi_{i,B}\}_{i=1}^{q}$ constructed such that the first $a$ and $b$ vectors of each set satisfy \eqref{eq:svdA} and \eqref{eq:svdB}, respectively. The remaining vectors are simply chosen to complete orthonormal bases for $\mathbb{R}^p$ and $\mathbb{R}^q$, respectively. Then build $\Psi_A$ as the matrix whose columns are the vectors of $\{\psi_{i,A}\}_{i=1}^p$ in the same order as in the set, and $\Psi_B$ as the matrix whose columns with indices from $a+1$ to $a+b$ are the first $b$ vectors of $\{\psi_{i,B}\}_{i=1}^{q}$ and whose remaining columns are the remaining vectors in the set in no particular order (here is where the assumption that $q>o$ is used, and the extra care on the order of the columns is necessary to make sure they match the order or the columns of $\Phi$).

            Finally, let $\Sigma_A\in\mathbb{R}^{p\times o}$ and $\Sigma_B\in\mathbb{R}^{q\times o}$ be rectangular diagonal matrices with the first $a$ elements of the main diagonal of $A$ being the elements of  $\{\sigma_{i,A}\}_{i=1}^{a}$, and the elements of the main diagonal of $\Sigma_B$ of indices from $a+1$ to $a+b$ being the elements of $\{\sigma_{i,B}\}_{i=1}^{b}$ (all remaining main diagonal elements are zero).
            
            With all matrices built as indicated, we can verify that
            \begin{equation*}
                \Psi_A\Sigma_A\Phi = \sum_{i=1}^a \psi_{i,A}\phi_{i,A}^\top\sigma_{i,A} = A
            \end{equation*}
            and
            \begin{equation*}
                \Psi_B\Sigma_B\Phi = \sum_{i=1}^b\psi_{i,B}\phi_{i,B}^\top \sigma_{i,B} = B,
            \end{equation*}
            which means that they are valid SVDs of $A$ and $B$, respectively. Furthermore, $\Sigma_A\Sigma_B^\top=0$ follows directly from their construction. \qed
            
            This concludes the proof of this auxiliary result, so we can proceed to proving the statement in Lemma \ref{lem:critcharact}.

            \textbf{Proof of Lemma \ref{lem:critcharact}:} To prove that $2)$$\Rightarrow$$ 1)$ we simply compute $[\dot K_1^\top ; \dot K_2]$ for a $(K_1, K_2)$ that satisfies the properties in $2)$ and verify that it is equal to zero. That is
            {
            \begin{align}
                    \begin{bmatrix}\dot K_1^\top \\ \dot K_2\end{bmatrix}=& \begin{bmatrix} 
                        \nabla_\K J(K_2K_1)^\top K_2 \\ \nabla_\K J(K_2K_1)K_1^\top 
                    \end{bmatrix} \nonumber\\=& \begin{bmatrix} 
                        \Phi{\Sigma}^\top\Psi^\top \Psi\Sigma_{ 2}\Gamma_{ 2}^\top \\ \Psi{\Sigma}\Phi^\top \Phi\Sigma_{ 1}\Gamma_{ 1}^\top
                    \end{bmatrix} \\=& \begin{bmatrix} 
                        \Phi{\Sigma}^\top\Sigma_{ 2}\Gamma_{ 2}^\top \\ \Psi{\Sigma}\Sigma_{ 1}\Gamma_{ 1}^\top 
                    \end{bmatrix} = \begin{bmatrix}
                        0 \\ 0
                    \end{bmatrix}.\nonumber
            \end{align}}
            
            To prove that $1)$$\Rightarrow$$2)$, apply Lemma \ref{lem:aux1} with $A = \nabla_\K J(K_2K_1)^\top$ and $B = K_2^\top$, which implies that $o=n\leq k=q$, since $\dot K_1 = -\nabla_{\K}J(K_2K_1)^\top K_2 = 0$, which allows us to write $ K_2$ and $\nabla_\K J(K_2K_1)$ as follows:
            {
            \begin{align}
                \label{eq:lmauxintoQ}
                    \nabla_\K J(K_2K_1) = \Psi \underbrace{\begin{bmatrix}
                        \bar \Sigma_{2} & 0 & 0 \\ 0 & 0 & 0 \\ 0 & 0 & 0
                    \end{bmatrix}}_{\Sigma_{2}} \Phi_{K_2}^\top \\ K_2 = \Psi \underbrace{\begin{bmatrix}
                        0 & 0 & 0 \\ 0 & \bar\Sigma_{K_2} & 0 \\ 0 & 0 & 0
                    \end{bmatrix}}_{\Sigma_{K_2}} \Gamma_2.\nonumber
            \end{align}}
            
            Similarly, applying Lemma \ref{lem:aux1} with $A = \nabla_\K J(K_2K_1)$ and $B = K_1$ gives 
            {
            \begin{align}
                \label{eq:lmauxintoP}
                    \nabla_\K J(K_2K_1) = \Psi_{K_1} \underbrace{\begin{bmatrix}
                        \bar \Sigma_{1} & 0 & 0 \\ 0 & 0 & 0 \\ 0 & 0 & 0
                    \end{bmatrix}}_{\Sigma_{1}} \Phi^\top \\ K_1 = \Gamma_{K_1} \underbrace{\begin{bmatrix}
                        0 & 0 & 0 \\ 0 & \bar\Sigma_{K_1} & 0 \\ 0 & 0 & 0
                    \end{bmatrix}}_{\Sigma_{K_1}}\Phi^\top.\nonumber
            \end{align}}
            
            Notice that even if $\bar\Sigma_{1}\neq\bar\Sigma_{2}$, they must still have the same diagonal elements, albeit possibly in a different order. Changing the order of the elements of $\bar\Sigma_{1}$ and $\bar\Sigma_{2}$ so they match means swapping the columns of $\Psi$, $\Phi$, $\Psi_{K_1}$ and $\Phi_{K_2}$, but we can also swap the singular vectors corresponding to the kernels of $K_2$ and $K_1^\top$ such that the results from Lemma \ref{lem:aux1} still hold. As such we can assume without loss of generality that $\bar\Sigma_{1}=\bar\Sigma_{2}=\bar\Sigma$. Notice that this is enough to prove that $1\rightarrow 2b$, since $\nabla_\K J(K_2K-1)K_1^\top=\Psi_{K_1}\Sigma\Phi^\top\Phi\Sigma_{K_1}^\top\Gamma_1=0\iff \Sigma\Sigma_{K_1}^\top=0$ (and similarly for $\Sigma^\top\Sigma_{K_2}=0$).
            
            Next we write
            {
            \begin{align}
                    \begin{bmatrix}
                    \Psi_{1,K_1} & \Psi_{2,K_1} & \Psi_{3,K_1}
                \end{bmatrix}&\begin{bmatrix}
                    \bar\Sigma & 0 & 0 \\ 0 & 0 & 0 \\ 0 & 0 & 0
                \end{bmatrix}\begin{bmatrix}
                    \Phi_1^\top \\ \Phi_2^\top \\ \Phi_3^\top
                \end{bmatrix} \\ = \begin{bmatrix}
                    \Psi_1 & \Psi_2 & \Psi_3
                \end{bmatrix}&\begin{bmatrix}
                    \bar\Sigma & 0 & 0 \\ 0 & 0 & 0 \\ 0 & 0 & 0
                \end{bmatrix}\begin{bmatrix}
                    \Phi_{1,K_2}^\top \\ \Phi_{2,K_2}^\top \\ \Phi_{3,K_2}^\top
                \end{bmatrix},\nonumber
            \end{align}}
            which implies that 
            {
            \begin{align}
                \label{eq:lemequality}
                \Psi_{1,K_1}\bar\Sigma\Phi_1^\top &= \Psi_1\bar\Sigma\Phi_{1,K_2}^\top.
            \end{align}}

            If we impose, for example, $\Psi_{1,K_1} = \Psi_1$, then we must also impose $\Phi = \Phi_{1,K_2}$. This leaves the SVD of $K_2$ intact, but changes part of the SVD of $K_1$. To show it still satisfies Lemma \ref{lem:aux1}, consider

            \begin{equation*}
                K_1^\top = \begin{bmatrix}
                    \Phi_{1,K_2} & \Phi_2 & \Phi_3
                \end{bmatrix}\begin{bmatrix}
                        0 & 0 & 0 \\ 0 & \bar\Sigma_{K_1} & 0 \\ 0 & 0 & 0
                    \end{bmatrix}\begin{bmatrix}
                    \Gamma_{1,P}^\top \\ \Gamma_{2,P}^\top \\ \Gamma_{3,P}^\top
                \end{bmatrix} = \Phi_2\bar\Sigma_{K_1}\Gamma_{2,P}^\top
            \end{equation*}
            which shows that the above is still a valid SVD of $K_1^\top$ as long as the span of the columns of $[\Phi_{1,K_2}, \Phi_3]$ is equal to the Kernel of $K_1$. Indeed, we know that $\colspan{[\Phi_{1}, \Phi_3]} = \kernel{K_1}$, and since \eqref{eq:lemequality} shows that $\colspan{\Phi_1} = \colspan{\Phi_{1,K_2}}$ (by contradiction) then we can conclude that $\colspan{[\Phi_{1,K_2}, \Phi_3]} = \kernel{K_1}$. We have, therefore, established that we can always match the singular vectors associated with the nonzero singular values of $\nabla_\K J(K_2K_1)$ for the SVDs in \eqref{eq:lmauxintoQ} and \eqref{eq:lmauxintoP} and still satisfy both conditions on Lemma \ref{lem:aux1}.

            Next, notice that we can simply pick $\Psi_{3,K_1} = \Psi_3$, $\Phi_{3,K_2} = \Phi_3$, since all matrices are related to the intersection of the kernels and their choice is arbitrary as long as they compose an orthonormal basis of $\kernel{\nabla_\K J(K_2K_1)}\cap\kernel{K_1}$ and of $\kernel{\nabla_\K J(K_2K_1)^\top}\cap\kernel{K_2^\top}$ respectively. 
            
            For the remaining matrices, $\Psi_2$ and $\Phi_2$ are imposed by the SVDs of $K_2$ and $K_1$ respectively, and as such cannot be changed arbitrarily. We can, however, freely change the columns of $\Psi_{2,K_1}$ (resp. $\Phi_{2,K_2}$) as long as when composed with the columns of $\Psi_3$ (resp. $\Phi_3$) they form a basis of the kernel of $\nabla_\K J(K_2K_1)^\top$ (resp. $\nabla_\K J(K_2K_1)$). Therefore we can select $\Psi_{2,K_1}$ (resp. $\Phi_{2,K_2}$) to be equal to $\Psi_2$ (resp $\Phi_2$) without any loss of generality, completing the proof.
        }

    \textbf{Proof of Corollary \ref{cor:finitesaddles}:}
        Let $(K_1,K_2)$ be a saddle point of the gradient flow dynamics \eqref{eq:GradFlowKi} with $N=2$ and assume $\rank{K_2K_1}=p<\min(m,n)$. Also, let $K^*$ be such that $\nabla_KJ(K^*)=0$. Then, let $v$ be any vector such that $v^\top \nabla_\K J(K_2K_1)=0$ and notice that
        \begin{align*}
            v^\top \nabla_KJ(K_2K_1) &= v^\top \nabla_KJ(K^*) \\
            v^\top R K_2K_1 &= v^\top RK^*.
        \end{align*}

        Then, let $[v_1,\dots,v_p]$ be $p$ linearly independent (LI) vectors such that $v_i^\top \nabla_KJ(K_2K_1)=0$, a set which must exist because from Lemma \ref{lem:critcharact} the left kernel of $\nabla_KJ(K_2K_1)$ has dimension $p$. Then, notice that for two LI vectors $u$ and $v$ and full rank matrix $R$, $u^\top R$ and $v^\top R$ must also be LI. Finally, let $\Psi_1^*$ be the matrix whose columns are vectors composing an orthonormal base of $\mbox{span}(v_1,\dots,v_p)$ and notice that
        \begin{equation*}
            (\Psi_1^*)^\top K_2K_1=(\Psi_1^*)^\top K^*,
        \end{equation*}
        however, since $K_2K_1$ is rank $p$, one can always pick the orthonormal basis that compose the columns of $\Psi_1^*$ to be the left singular vectors of $K_2K_1$, implying that there exist a $\Phi_1^*$ whose columns are orthonormal vectors such that
        \begin{equation*}
            (\Psi_1^*)^\top K_2K_1 \Phi_1^* = \Sigma_1^* = (\Psi_1^*)^\top K^* \Phi_1^*.
        \end{equation*}
        
        For the remaining components of the SVD of $K^*$ we can pick whichever eigenvectors are left since they are all a basis for the left kernel of $K_2K_1$. \qed
    
\section{Proofs for the Simple Example}         


        \textbf{Proof of Proposition \ref{prop:convergence}:} In this proof it is assumed that $a<0$ since if $a>0$ then $(k_1,k_2)=(0,0)\not\in\mathcal{K}$ and any solution initialized in $\mathcal{K}$ always converges to $\mathcal{T}:=\{(k_1,k_2)\in\mathcal{K} ~|~ k_2k_1=k^*\}$. 
        
        We first point out a natural division of the state space in the invariant sets $\{(k_1,k_2)\in\mathcal{K} ~|~ f(k_1,k_2)>0\}$ and $\{(k_1,k_2)\in\mathcal{K} ~|~ f(k_1,k_2)<0\}$, since any solution initialized such that $f(k_1,k_2)>0$ would need to cross $\mathcal{T}$ before reaching a point such that $f(k_1,k_2)<0$ (and vice versa). 
        
        Furthermore, notice that $f(0,0)=-q/2a>0$ since $a<0$ by assumption (argued above). So for any initialization such that $f(k_1,k_2)<0$, the only reachable critical points are in $\mathcal{T}$. 

        For initializations such that $f(k_1,k_2)>0$, consider the following change of coordinates
        \begin{align*}
            \ell_1 &= (k_1+k_2^\top)0.5 \\
            \ell_2 &= (k_1-k_2^\top)0.5,
        \end{align*}
        which induces the following dynamics
        \begin{align*}
            \dot\ell_1 = -f(\ell_1,\ell_2)\ell_1 \\
            \dot \ell_2 = f(\ell_1,\ell_2)\ell_2.
        \end{align*}
        Notice that for $f(\ell_1,\ell_2)\geq 0$, $\|\ell_1(t)\|_2^2$ is non-increasing and $\|\ell_2(t)\|_2^2$ is non-decreasing. This means that if $\ell_2(0)\neq0$, $\ell_2(t)\neq 0$ for all $t>0$. Since all nonzero critical points lie in $\mathcal{T}$ and all solutions converge to a critical point (by Theorem \ref{thm:existence}) then $\|\ell_2(0)\|_2^2=\|k_1-k_2^\top\|_2^2>0$ is sufficient for convergence to the target set. 
        
        To show that it is necessary, first notice that if $k_1=k_2^\top=k$ then $k_2k_1=k^\top k\geq 0>k^*$, therefore any point such that $\ell_2=0$ lies in the part of the state space such that $f(k_1,k_2)=f(\ell_1,\ell_2)>0$. For an initialization such that $\ell_2=0$, $\dot\ell_2=0$ for all time, therefore it is such that $\ell_2(t)=0$. The only finite critical point that is such that $\ell_2=0$ is the spurious equilibria $(k_1,k_2)=0$, showing that $\|\ell_2(0)\|_2^2>0$ is necessary for convergence to the target set, and completing the proof. \qed
%
    
        \textbf{Proof of Proposition \ref{prop:fastconv}:} Compute the ODE associated with $J(k_1,k_2)$ as 
        \begin{align}
            \dot J &= -f(k_1,k_2)(k_2\dot k_1 + \dot k_2k_1) \nonumber\\&= -f(k_1,k_2)^2(k_1^\top k_1+k_2k_2^\top) \nonumber\\&= -f(k_1,k_2)^2(\trace{\mathcal{C}}+2k_2k_2^\top) \label{eq:Jdot}
        \end{align}
        Since by assumption, the proposition compares two points with the same value for the cost function, all we need to evaluate form \eqref{eq:Jdot} is the value of $\trace{\mathcal{C}}+2k_2k_2^\top$. To do that, square and take the trace from both sides of \eqref{eq:invexp}, resulting in
        \begin{equation*}
            \trace{\mathcal{C}^2} = (k_1^\top k_1)^2-2(k_2k_1)^2+(k_2k_2^\top)^2.
        \end{equation*}
        Using the fact that $k_2k_1=k$ is constant along all initializations and that $k_1^\top k_1 = \trace{\mathcal{C}}+k_2k_2^\top$ we get
        \begin{align*}
            (k_2k_2^\top)^2&+\trace{\mathcal{C}}k_2k_2^\top \\&+\frac{\trace{\mathcal{C}}^2-\trace{\mathcal{C}^2}-2k^2}{2}=0,
        \end{align*}
        which has, as only real solution:
        \begin{align*}
            2k_2k_2^\top &= -\trace{\mathcal{C}}\\&+\sqrt{2\trace{\mathcal{C}^2}-\trace{\mathcal{C}}^2+4k^2},
        \end{align*}
        which in turn implies that
        \begin{equation*}
            \trace{\mathcal{C}}+2k_2k_2^\top = \sqrt{c+4k^2}
        \end{equation*}
        for $c:=2\trace{\mathcal{C}^2}-\trace{\mathcal{C}}^2$, as defined Definition \ref{def:vecimbalance}. This expression is strictly increasing in $c$, showing that for any two points $(\tilde k_1, \tilde k_2)$ and $(\overline k_1, \overline k_2)$ such that $\tilde c > \bar c$, $\dot J(\tilde k_1, \tilde k_2)<\dot J(\overline k_1, \overline k_2)\leq0$. 
        
        Let $\alpha(t) = J(\overline k_1(t),\overline k_2(t))-J(\tilde k_1(t),\tilde k_2(t))$. To conclude this proof, it is enough to show that $\alpha>0$ for all $t>0$. To see that, consider that by construction, $\alpha(t)=0$ and $\dot\alpha|_{t=0}>0$, which means $\exists \epsilon>0$ such that $\alpha(\epsilon)>0$. Assume there exists some $\bar t>0$ such that $\alpha(\bar t)=0$. Furthermore, assume without loss of generality that $\bar t$ is minimal (\ie the first $t>0$ for which this condition holds). Then, from the previous result, one can conclude that $\dot J(\tilde k_1(\bar t), \tilde k_2(\bar t))<\dot J(\overline k_1(\bar t), \overline k_2(\bar t))$ which means that $\dot \alpha|_{t=\bar t}>0$, implying that there exists $\delta>0$ such that $\alpha(\bar t-\delta)<0$. However, by the mean value theorem, there must exist $\tilde t\in(0,\bar t-\delta)$ such that $\alpha(\tilde t)=0$, reaching contradiction. Therefore, $\alpha(t)>0$ for all $t>0$. \qed
        




        \textbf{Proof of Proposition \ref{prop:ISS}:} To begin the proof of Proposition \ref{prop:ISS} we first prove the following auxiliary result

        \begin{proposition}
            Let $\mathcal{R}_\alpha:\{(k_1,k_2)\in\mathcal{K} ~|~ \|k_1-k_2^\top\|_2^2\geq \alpha^2\}$. Then, for $\alpha\in(0,2\sqrt{|k^*|})$, $R_\alpha$ is forward-invariant under the disturbed gradient-flow from \eqref{eq:k1dotdist} if
            \begin{equation*}
                \|u_1-u_2^\top\|_\infty \leq \alpha\frac{r\alpha^4-8ar\alpha^2-16q}{2(4a-\alpha^2)}
            \end{equation*}
        \end{proposition}

            To prove this result, let the point $(\overline k_1,\overline k_2)$ be the value of a trajectory at time $t=\bar t$ at the border of $\mathcal{R}_\alpha$, that is $\|\overline k_1-\overline k_2^\top\|_2^2=\alpha^2$. It holds that
            \begin{equation*}
                \frac{d}{dt}\|\overline k_1-\overline k_2^\top\|_2^2>0.
            \end{equation*}
            Computing the time derivative gives
            \begin{align*}
                \frac{d}{dt}\|\overline k_1-\overline k_2^\top\|_2^2&= 2(\overline k_1-\overline k_2^\top)^\top (\dot k_1 -\dot k_2^\top ) \\ &= 2f(\overline k_1,\overline k_2)(\overline k_1-\overline k_2)^\top(\overline k_1-\overline k_2)\\&~~-2(\overline k_1-\overline k_2)^\top(u_1(\bar t)-u_2(\bar t)^\top) \\ &\geq 2f(\overline k_1,\overline k_2)\|\overline k_1-\overline k_2^\top\|_2^2\\&~~-2\|\overline k_1-\overline k_2^\top \|_2\|u_1(\bar t)-u_2(\bar t)^\top \|_2.
            \end{align*}
            Notice that the lower bound above is greater than zero if 
            \begin{equation*}
                \|u_1-u_2^\top\|_\infty\leq f(\overline k_1,\overline k_2)\|\overline k_1-\overline k_2^\top \|_2.
            \end{equation*}
            To lower bound the RHS of the expression above by a function of $\alpha$, consider the expression for $f(k_1,k_2) := f(k_2k_1)$ (notice the notation overload)
            \begin{equation*}
                f(k_2k_1) = -\frac{r(k_2k_1)^2+2ark_2k_1-q}{2(a+k_2k_1)^2}.
            \end{equation*}
%
            Notice that $f(k_2k_1)\rightarrow+\infty$ as $k_2k_1\rightarrow -a$. Furthermore, taking the gradient of $f(k_2k_1)$ gives
            \begin{equation*}
                \frac{\partial}{\partial (k_2k_1)}f(k_2k_1) = -\frac{ra^2+q}{(a+k_2k_1)^3}
            \end{equation*}
            which is positive for all $k_2k_1<-a$. This means that for $k_2k_1<-a$, $f(k_2k_1)$ is monotonically increasing, going to zero as $k_2k_1\rightarrow -\infty$ and to $+\infty$ as $k_2k_1\rightarrow -a^-$. Furthermore, from $\|k_1-k_2^\top \|_2^2=\alpha^2$ one can easily derive that $k_2k_1\geq -\alpha^2/4$ which in turn means that $f(k_2k_1)>f(-\alpha^2/4)$. We can, then, rewrite the bound on the infinity norm of $u_1$ and $u_2$ as
            \begin{align*}
                \|u_1-u_2\|_\infty^\top &\leq f(-\alpha^2/4)\alpha \\ &= \alpha\frac{r\alpha^4-8ar\alpha^2-16q}{2(4a-\alpha^2)}
            \end{align*}
            which recovers the sufficient condition in the statement.
            \begin{remark}
                To see that the bound above is not empty, first look at
                \begin{equation*}
                    \mathcal{R}(\alpha) = -\alpha\frac{r\alpha^4 - 8ar\alpha^2 - 16q}{2(-\alpha^2 + 4a)^2},
                \end{equation*}
                and consider the polynomial in $\alpha$ taken from the numerator of $\mathcal{R}(\alpha)$:
                \begin{equation*}
                    \mathcal{P}(\alpha) = -\alpha(r\alpha^4 - 8ar\alpha^2 - 16q),
                \end{equation*}
                which has five roots, denoted as follows: 
                \begin{align*}
                    \alpha_{1,2}&=\pm 2\sqrt{a+\sqrt{a^2+q/r}}\in\mathbb{R}\\
                    \alpha_{3,4}&=\pm 2i\sqrt{-a+\sqrt{a^2+q/r}}\in\mathbb{I}\\
                    \alpha_0&=0\in\mathbb{R}.
                \end{align*}
                The interval of interest for the analysis is inside (sometimes equal to) the interval between $\alpha = 0$, where the line $|k_1-k_2|=0$ contains the point $k_1=k_2=0$ which is a spurious equilibrium of the system, and $\alpha=2\sqrt{|k^*_-|} = \alpha_1$, where the line $|k_1-k_2|=2\sqrt{|k^*_-|}$ contains the point $k_1=-k_2=\sqrt{|k^*_-|}$ which is part of the target equilibrium set of the system. In other words, if $\alpha>0$ it is guaranteed that the set $|k_1-k_2|>\alpha$ does not contain the spurious equilibrium at the origin, and if $\alpha<2\sqrt{|k^*_-|}$, the set $|k_1-k_2|>\alpha$ contains the entirety of the target set $k_2k_1=k^*_-$. {These two lines are indicated in Fig. \ref{fig:PPScalarCase} by the green and blue lines, respectively for the scalar case.} 
                
                To evaluate the sign of $\mathcal{P}(\alpha)$ between $\alpha_0$ and $\alpha_1$, evaluate its derivative at $\alpha_0$
                \begin{align*}
                    \mathcal{P}'(\alpha_0)&= -(r\alpha^4 - 8ar\alpha^2 - 16q)-\alpha(4r\alpha^3 - 16ar\alpha) \\
                    &= 16q >0.
                \end{align*}
                Therefore, for $\alpha\in (\alpha_0,\alpha_1) = (0,2\sqrt{|k^*_-|})$, $\mathcal{P}(\alpha)>0$ which means that the bound $\|u-v\|_\infty<\mathcal{B}_{uv}(k_2)$ is not empty has a maximum value inside $(0,2\sqrt{|k^*_-|})$.
            \end{remark}

        Next we prove Proposition \ref{prop:ISS}. Using the fact that $\mathcal{R}_\alpha$ is invariant, one can prove Proposition \ref{prop:ISS} as follows: First, consider $J(k_1,k_2)$ as a candidate Lyapunov Function of the gradient dynamics. Then, let $U=[u_1;u_2^\top]$ and compute
        {\small
        \begin{align*}
            \dot J &(k_1,k_2) = \langle\nabla J,-\nabla J + U\rangle \\
            &= -\|\nabla J\|^2+\langle\nabla J,U\rangle \\
            &\leq -\|\nabla J\|^2+\|\nabla J\|\|U\| \\
            &\leq -\frac{1}{2}\|\nabla J\|^2+\frac{1}{2}\|U\|^2\\
            &=-\frac{1}{2}\big(f(k_1,k_2)^2(\|k_1\|_2^2+\|k_2\|_2^2)\\
            &~~~~+\|u_1\|^2+\|u_2\|^2\big) \\
            &\leq -\frac{1}{2}\left(\frac{r(k_2k_1)^2+2ark_2k_1-q}{2(a+k_2k_1)^2}\right)^2(\|k_1\|_2^2+\|k_2\|_2^2) \\ &~~~~+\frac{1}{2}(\|u_1\|_\infty^2+\|u_2\|_\infty^2)\\
            &= -r\frac{(k^*_+-k_2k_1)^2(k^*_--k_2k_1)^2}{2(a+k_2k_1)^4}(\|k_1\|_2^2+\|k_2\|_2^2)\\ &~~~~+\frac{1}{2}(\|u_1\|_\infty^2+\|u_2\|_\infty^2),
        \end{align*}
        }
        where $k_+^*$ and $k_-^*$ are defined as in \eqref{eq:Zerosfp} and \eqref{eq:Zerosfm} respectively.

        Next, notice that for a given $\epsilon\geq0$, if $J(k_1,k_2)\geq \tilde\epsilon=\epsilon+J^*$ then (noticing that $(a+k2k1)<0$ by assumption)
        \begin{align*}
            &-\frac{(q+(k_2k_1)^2r)}{2(a+k_2k_1)}\geq \tilde\epsilon \\
            & (k_2k_1)^2r+2\tilde\epsilon k_2k_1+2a\tilde\epsilon+q\geq 0
        \end{align*}
        which holds if ant only if 
        \begin{align*}
            &k_2k_1\leq-\tilde\epsilon/r-\sqrt{\tilde\epsilon^2-2ar\tilde\epsilon-rq}/r, \text{~or} \\&k_2k_1\geq-\tilde\epsilon/r+\sqrt{\tilde\epsilon^2-2ar\tilde\epsilon-rq}/r.
        \end{align*}

        Notice that this interval is always nonempty for $\epsilon>0$ since $\tilde\epsilon^2-2ar\tilde\epsilon-rq$ has its zeros at $\tilde\epsilon=ar\pm r\sqrt{a^2+q/r}$, and $\tilde\epsilon \geq J^*=ar+ r\sqrt{a^2+q/r}$.

        Also notice that for $\epsilon\rightarrow\infty$ the intervals simplify to $k_2k_1\leq-\infty$ (which matches the coerciveness of the cost function) or $k_2k_1\geq-a$ (which matches the fact that the cost explodes at the border of instability).


        With this estabilished, now consider $|k^*_--k_2k_1|$ as a measure of the size of the state, and notice from the previous argument that for every $\epsilon>0$ there exists a $\gamma>0$ such that if $J(k_1,k_2)\geq\epsilon$ then $|k^*_--k_2k_1|\geq \gamma$. Then notice that for $k=k_2k_1<-a$, the rational function $h:\mathbb{R}\rightarrow\mathbb{R}$ defined as $h(k) = \frac{(k^*_+-k)^2}{(a+k)^4}$ has no inflection points, since 
        \begin{equation*}
            \frac{\partial}{\partial k}h(k) = -4\frac{(k-k_+^*)((a+2k_+^*)-k)}{(a+k)^5}
        \end{equation*}
        is zero only if $k=k_+^*\not\in\mathcal{K}$ or if $k=(a+2k_+^*)\not\in\mathcal{K}$.
        
        Furthermore, it goes to infinity as $k_2k_1\rightarrow -a^-$ and to zero as $k_2k_1\rightarrow -\infty$. This means that in the interval of interest given by $k^*_-\leq k_2k_1<-a$, the rational function $h(k_2k_1)$ is lower bounded by $h(k^*_-) = 4/(a^2+q/r)$.

        Furthermore, since we established that $\mathcal{R}_\alpha$ is invariant for sufficiently bound disturbances, assuming there exist small enough $\alpha>0$ such that the solution is initialized inside $\mathcal{R}_\alpha$, one can bound $\|k_1\|_2^2+\|k_2\|_2^2\geq \alpha^2/2$, which in turns allows one to write
        \begin{align*}
            \dot J &\leq -r\frac{2}{a^2+q/r}\frac{\alpha^2}{2}(k_-^*-k_2k_1)^2\\&~~~~+\frac{1}{2}(\|u\|_\infty^2+\|v\|_\infty^2)\\
            &\leq -r\frac{2}{a^2+q/r}\frac{\alpha^2}{2}\gamma^2\\&~~~~+\frac{1}{2}(\|u\|_\infty^2+\|v\|_\infty^2)
        \end{align*}
        From this, one notice that $$\delta^2=\min\left(r\frac{2}{a^2+q/r}{\alpha^2}\gamma^2,0.5(f(-\alpha^2/4)\alpha)^2\right)$$ is such that if $\|u_1\|_\infty^2+\|u_2\|_\infty^2\leq \delta^2$ then $\|u_1-u_2^\top\|^2_\infty\leq 2\|u_1\|_\infty^2+2\|u_2\|_\infty^2 \leq f(-\alpha^2/4)\alpha$, and $\dot J<0$ for all $k_2k_1$ such that $J(k_1,k_2)-J^*>\epsilon$, proving the statement of the proposition. \qed

\end{document}